\documentclass{aa}
\usepackage{graphicx}
\usepackage{longtable}
\usepackage{latexsym}
\usepackage{amssymb}
\usepackage{lscape}
\usepackage{natbib}
\begin{document}

\title{Witnessing galaxy preprocessing in the local Universe: 
the case of a star-bursting group falling into Abell 1367\thanks{Based
on observations collected at the European Southern Observatory
(proposal 71B-0154B), at the Italian Telescopio Nazionale Galileo (TNG) operated
on the island of La Palma by the Centro Galileo Galilei of
the INAF (Istituto Nazionale di Astrofisica) at the
Spanish Observatorio del Roque de los Muchachos of the Instituto de Astrofisica de Canarias,
at the Observatoire de Haute
Provence (OHP) (France), operated by the CNRS, 
and at the Loiano
telescope belonging to the University of Bologna (Italy).}}
\author{L. Cortese\inst{1,2,3}, G. Gavazzi\inst{1}, A. Boselli\inst{2}, P. Franzetti\inst{4}, R. C. Kennicutt\inst{5}, K. O'Neil\inst{6}, S. Sakai\inst{7}}
  \institute{Universit\'{a} degli Studi di Milano-Bicocca, P.zza della Scienza 3, 20126 Milano, Italy.
 	 \and 
 	     Laboratoire d'Astrophysique de Marseille, BP8, Traverse du Siphon, F-13376 Marseille, France.
	 \and
	     Present address: School of Physics and Astronomy, Cardiff University, Cardiff CF24 3YB, UK.\\
	     \email{luca.cortese@astro.cf.ac.uk}
	 \and     
	     IASF-INAF, Sezione di Milano, Via E. Bassini 15, I-20133 Milano, Italy
	 \and
	     Institute of Astronomy, University of Cambridge, Madingley Road, Cambridge. CB3 0HA,UK.
	 \and
	     NRAO, P.O. Box 2, Green Bank, WV 24944     
	 \and
	     Department of Astronomy, University of California, Los Angeles,  CA, 90095-1562
 	     }

   \date{Received 17 January 2006| Accepted 30 March 2006}

\abstract{
We present a multiwavelength analysis of a compact group of galaxies infalling 
at high speed into the dynamically young cluster Abell 1367.
Peculiar morphologies and unusually high H$\alpha$ emission are 
associated with two giant galaxies and at least ten dwarfs/extragalactic HII regions, making this
group the region with the highest density of star formation activity ever observed in the local clusters.
Moreover H$\alpha$ imaging observations reveal extraordinary complex trails of ionized gas behind 
the galaxies, with projected lengths exceeding 150 kpc. These unique 
cometary trails mark the gaseous trajectory of galaxies, witnessing 
their dive into the hot cluster intergalactic medium.
Under the combined action of tidal forces among group members and the ram-pressure by the 
cluster ambient medium, the group galaxies were fragmented and the ionized gas was blown out. 
The properties of this group suggest that environmental effects within infalling groups may have represented 
a \emph{preprocessing} step of the galaxy evolution during the high redshift cluster assembly phase. 

}

\titlerunning{Galaxy preprocessing in the local Universe.}
\authorrunning{L. Cortese et al.}

   \maketitle
%

\section{Introduction}
Cluster of galaxies are unique laboratories 
for studying the dependence of galaxy evolution on the local environment.
In the last decades a plethora of observational evidence has been collected, 
showing that the cluster environment is capable of perturbing both the morphology and 
the star formation history of galaxies (i.e. \citealp{GUNG72,dressler80,harrassment,poggianti01,gav02,n4438}).
Most of what is known about the evolution of galaxies in clusters 
derives from studies of cluster cores, where the environmental effects are more 
severe, both locally 
and at higher redshift (i.e. \citealp{cayatte90,smail97,hector1,n4438}).
Galaxies, however, interact with the harsh environment 
well before reaching the center of a cluster \citep{kodama01,treu03}.
Moreover, the galaxies likely belong to infalling groups of galaxies which, according 
to the hierarchical scenario of the formation of large scale structures, 
represent the building blocks of today rich clusters of galaxies.    
Galaxy groups may therefore represent natural sites for a 
\emph{preprocessing} stage in the evolution of cluster galaxies \citep{FUJI04,mihos04}
through tidal interactions, otherwise ineffective in high velocity 
dispersion environments.
This suggests that at least part of the morphological transformation 
of cluster galaxies took place in earlier epochs 
in very different conditions than the ones observed in today's clusters \citep{dress04car}.
Unfortunately, witnessing \emph{preprocessing} in local Universe 
is a real challenge 
since the current merger rate is considerably lower than in the past \citep{lambcdm}. 
Presently we can observe several clusters experiencing 
multiple merging \citep{gav99,donnelly01,ferrari03,COGA04}, but the structures involved 
are sub-clusters with a mass $\sim5\times10^{14} \rm ~M_{\odot}$, considerably higher 
than the typical mass of a compact group $\sim10^{13} \rm ~M_{\odot}$ \citep{cgroup}.\\
This is also the case for the Abell cluster 1367  (z$\sim$0.0216) which is currently 
forming at the intersection of two filaments in the Great Wall,
as revealed by the recent dynamical study of \cite{COGA04}. 
Abell 1367 is a unique example among local, dynamically 
young clusters since, in addition to massive evolved substructures, it is also experiencing the merging 
of a compact group directly infalling into its core \citep{COGA04}.
This merger was independently discovered by \cite{jorge} 
and \cite{SAKK02} during deep H$\alpha$ surveys 
of nearby clusters, regions with the highest density of star forming systems 
ever observed in the local Universe.
We refer to this infalling group as the Blue Infalling Group (BIG), as named by \cite{GAVC03}.
BIG has a mean recessional velocity of $<V>=8230~ \rm km~s^{-1}$ and a dispersion of only 
$\sim150\rm km~s^{-1}$ \citep{GAVC03}.
\cite{SAKK02} argued that BIG lies in the cluster background, since it exceeds by $\sim1700\rm km~s^{-1}$ 
the mean cluster velocity of $<V>=6484~ \rm km~s^{-1}$ ($\sigma_V=891~\rm km~s^{-1}$, \citealp{COGA04}). 
Contrary to this, the dynamical analysis presented by \cite{COGA04}, is consistent 
with an infalling scenario, as also proposed by \cite{GAVC03}. 
Moreover this picture of BIG as an infalling cluster is supported by X-ray observations -- 
\cite{SUNM02}, using Chandra observations, discovered extended gas features and a
ridge near the SE cluster center. They proposed that these features are associated 
with a new merging component penetrating the SE sub-cluster.
These observational evidences suggest that we are witnessing, to our knowledge 
for the first time in the local Universe,
a compact group infalling into a core of a dynamically young cluster.
Because of its proximity, it represents a unique laboratory, allowing us to study in great detail
a physical process that should be typical of clusters at higher redshift. 
The study of this group could help us shed light 
on the role of \emph{preprocessing} on the evolution of galaxies in today's rich clusters.\\
During the last few years we have collected a multiwavelength dataset aimed at 
reconstructing the evolutionary history of this unique group of galaxies.
The observations and data reduction are presented in Sect.2; 
 the peculiar properties of this group are described in 
Sect.3 and their origin is discussed in Sect.4, including a comparison with the Stephan's Quintet.
Finally in Sect.5 we conclude with a possible evolutionary scenario for this group
and we discuss the importance of infalling groups on the evolution of cluster galaxies at higher redshift. 
Throughout this paper we assume for Abell 1367 a distance modulus $\rm \mu=34.84$ mag \citep{sakai00},
corresponding to $\rm H_{0}= 70~ km~s^{-1}~Mpc^{-1}$.

\section{Observations}
\subsection{HI observations}
\begin{figure*}
\centering
\includegraphics[width=8.5cm]{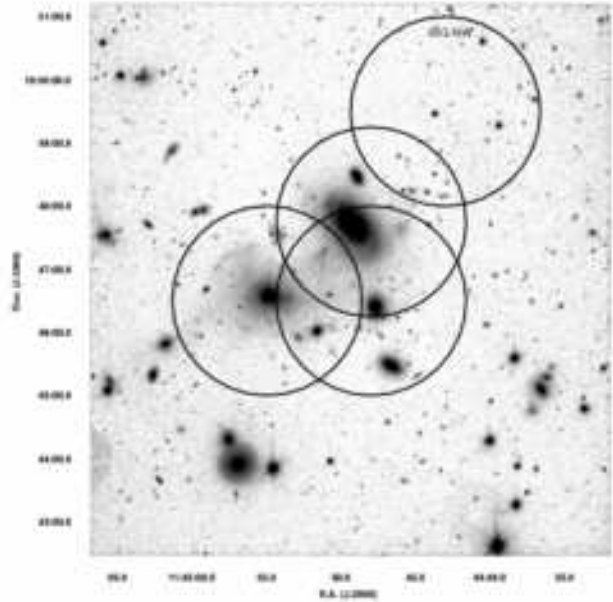}
\small{\caption{The four Arecibo HI pointings obtained in the region of the BIG group, 
superposed to the $r'$ band image. 
The size of each circle correspond to the telescope beam.}
\label{HIpoints}}
\end{figure*}

\begin{figure*}
\centering
\includegraphics[width=10.cm]{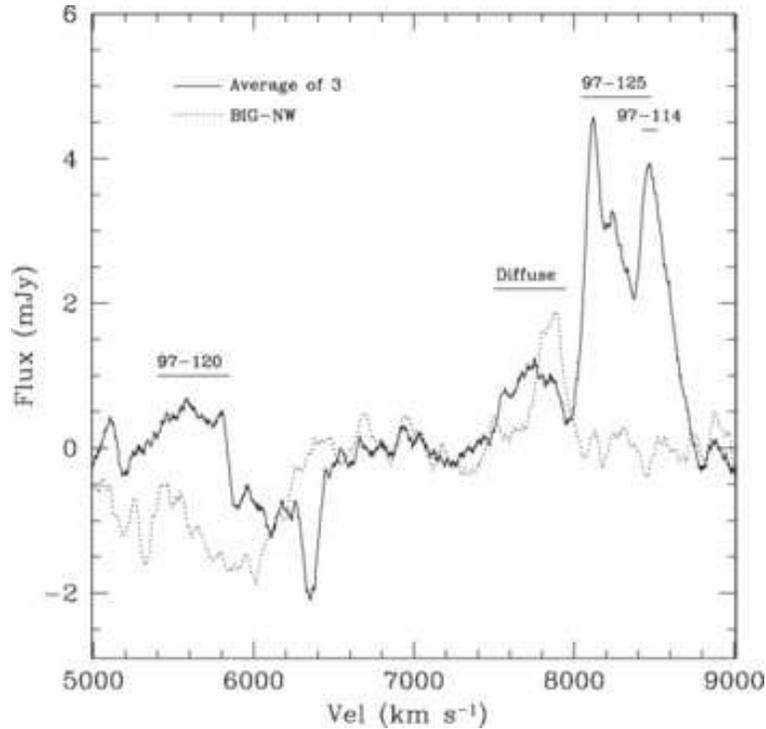} 
\small{\caption{
Comparison between the combined HI spectrum obtained from the three Arecibo pointings 
(97-125, 97-120, 97-114; solid line) 
and the independent BIG-NW pointing (dotted line). The optical velocity width of the three bright Zwicky 
galaxies is given. The diffuse HI component is also indicated.
}\label{HI2}}
\end{figure*}

Using the refurbished 305-m Arecibo Gregorian radio telescope we observed  the 
BIG region at 21 cm in March 2005. We obtained observations in 4 positions covering 
the group center and its NW outskirt (see Fig.\ref{HIpoints}). Given the instrument's beam at 21 cm 
(3.5'$\times$3.1') the three central positions (97-114,97-120 and 97-125), unlike BIG-NW,  
are not independent.
Data were taken with the L-Band Wide receiver, using nine-level sampling with two of 
the 2048 lag subcorrelators set to each polarization channel. All observations were taken 
using the position-switching technique, with each blank sky (or OFF) position observed for 
the same duration, and over the same portion of the telescope dish (Az and El) as the 
on-source (ON) observation. Each 5min+5min ON+OFF pair was followed by a 10s ON+OFF 
observation of a well-calibrated noise diode. Three pairs of observations were obtained per position, reaching 
an RMS noise of $\sim$0.2 mJy.
The velocity resolution was 2.6 $\rm km~s^{-1}$ and the pointing accuracy is about 15$''$. 
Flux density calibration corrections are good to within 10\% or better \citep{oneil04}.\\  
Using standard IDL data reduction software available at Arecibo, corrections were applied 
for the variations in the gain and system temperature with zenith angle and azimuth. A 
baseline of order one to three was fitted to the data, excluding those velocity ranges 
with HI line emission or radio frequency interference (RFI). The velocities were corrected 
to the heliocentric system, using the optical convention, and the polarizations were 
averaged. All data were boxcar smoothed to a velocity resolution of 12.9 $\rm km~s^{-1}$ for further 
analysis. The obtained spectra are given in Fig.\ref{HI2}.

\subsection{H$\alpha$ imaging}
\begin{figure*}
\centering
\includegraphics[width=19cm]{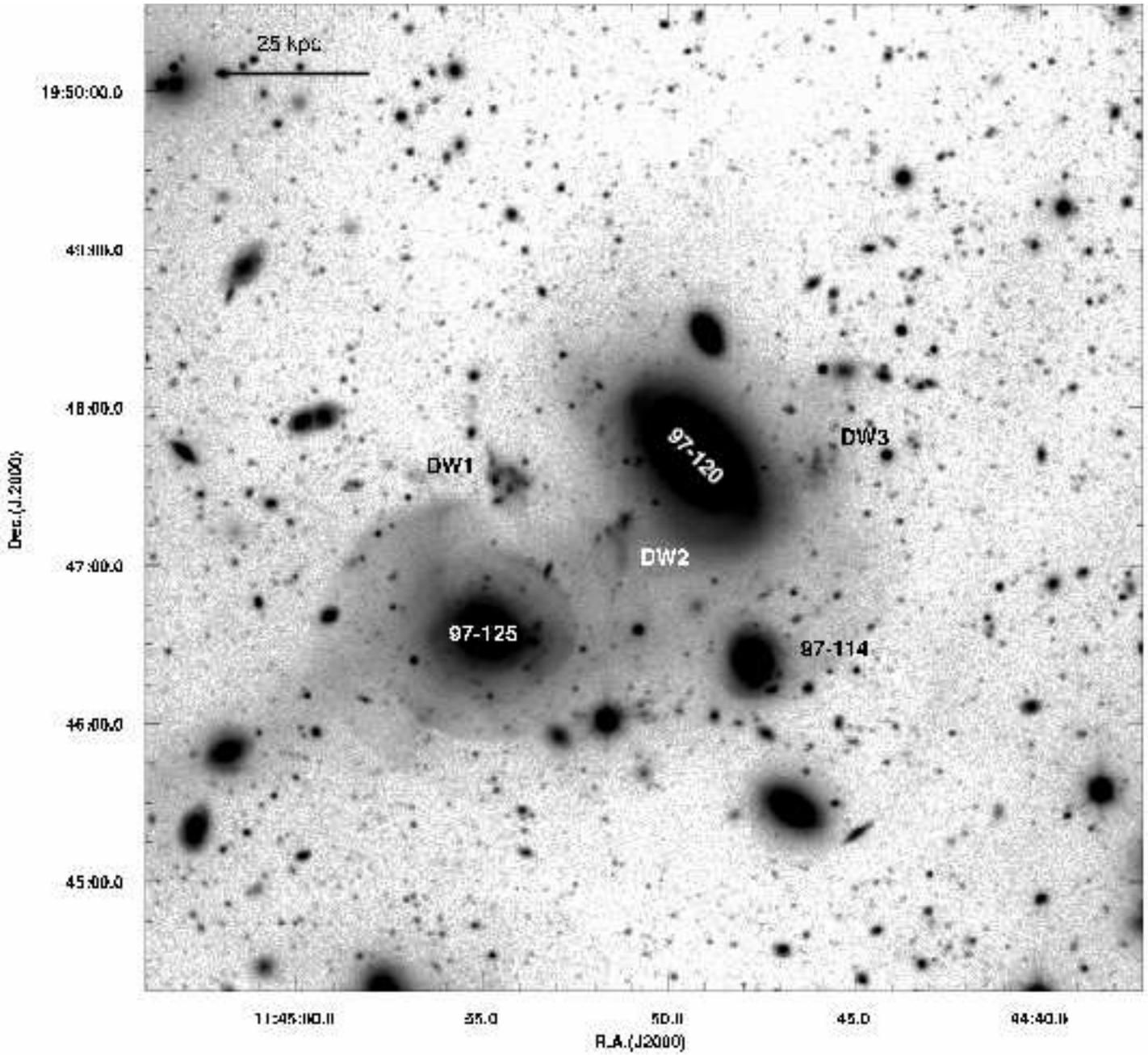}
\small{\caption{Stellar shells are seen around galaxy CGCG97-125 in the 
$r'$ band image of BIG. The three bright galaxies and the three dwarfs composing BIG are labeled.}
\label{R}}
\end{figure*}

We observed BIG using the Device Optimized for the LOw RESolution (DOLORES)
attached at the Nasmyth B focus of the 3.6m TNG in the photometric nights of 17th 
May and 18th June, 2004. The observations were
taken through a $\rm [SII]$ narrow band filter centered at $\rm
\sim6724\AA$ with a width of $\rm \sim57 \AA$ covering the
red-shifted H$\alpha$ and $\rm[NII]$ lines. The underlying
continuum was taken through a broadband (Gunn) $r'$ filter. 
Images were split in 6 exposures of
1200 sec in the narrow band filter and 5 exposures of 300
sec in the $r'$ broadband filter, for a total of 2
hours and 25 minutes exposure respectively.
Seeing was $\sim 1.2$ arcsec.
Photometric calibration was
achieved by observing the spectrophotometric star Feige 34. After
bias subtraction and flat-fielding, the images were combined. The
intensity in the combined OFF-band frame was normalized to that of
the combined ON-band one by the flux ratio of several field
stars. The NET image was obtained by subtracting the
normalized OFF-band frame to the ON-band one. 
H$\alpha$+[NII] fluxes and EWs are obtained as described 
in \cite{bosha2}.\\
Ten additional $r'$ exposures of 600 sec each were taken in the photometric 
nights of 1st-4th February 2006 using the same telescope, instrumental set-up and 
with a mean seeing of $\sim1.1$ arcsec.
The resulting image, obtained from the combination of all the $r'$ exposures, has a total 
exposure time of 125 minutes, reaching a 1$\sigma$ depth of 
$\mu_{r'}=27.4~ \rm mag ~arcsec^{-2}$.
The $r'$-band and NET frames are shown in Figs.
\ref{R} and \ref{net} respectively.

\subsection{MOS spectroscopy}
\label{datareduct}
\begin{figure*}
\centering
\includegraphics[width=18cm]{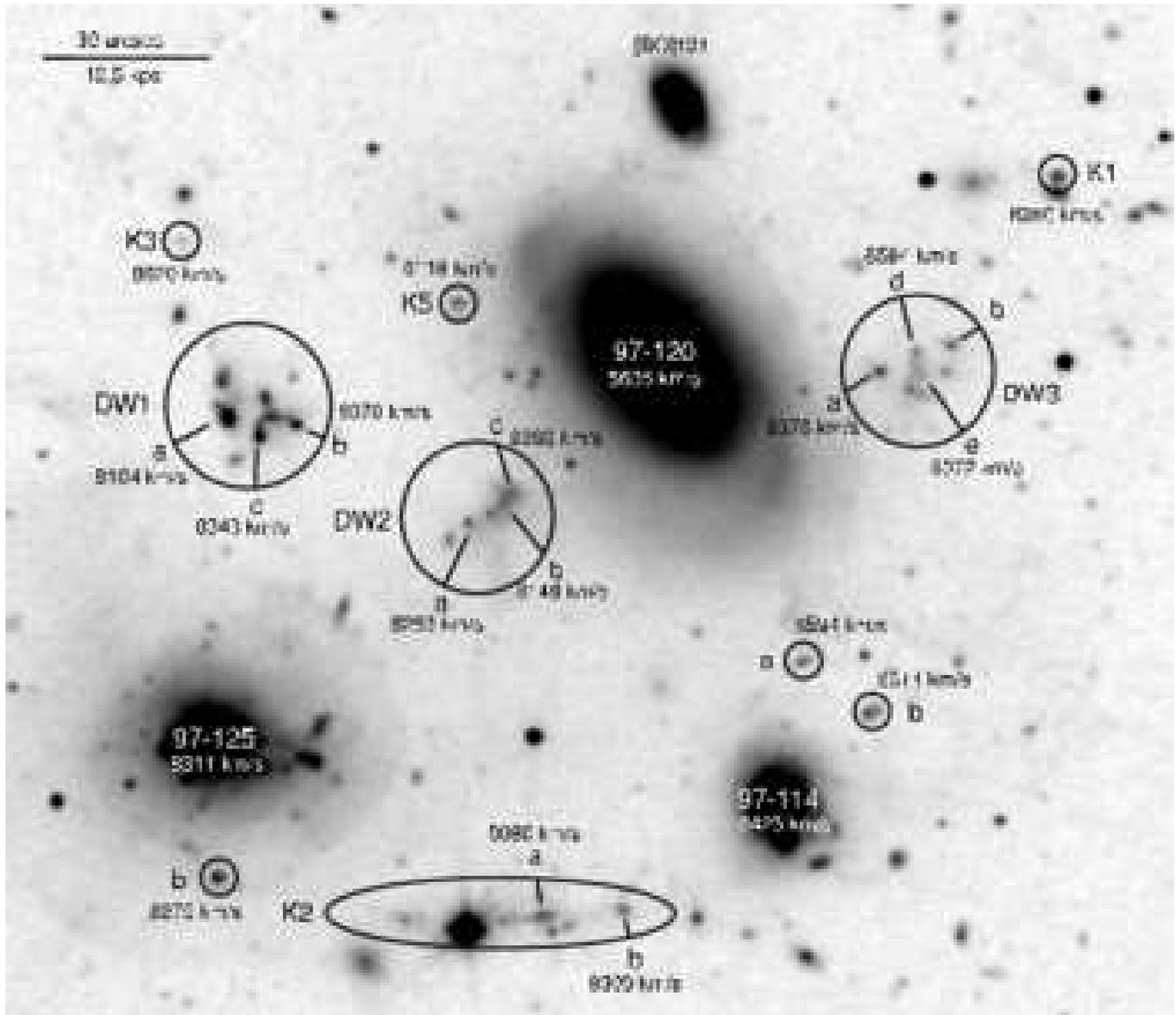} 
\small{\caption{
High-contrast H$\alpha$+[NII] (ON) band frame of the BIG group. 
Labels indicate the star forming regions and their associated velocities.}\label{ON}}
\end{figure*}

We observed the BIG region in MOS mode with the ESO/3.6m and with the 
TNG telescope. The ESO/3.6m observations were taken in the
photometric nights of May 5th and 6th 2003 with the ESO Faint
Object Spectrograph and Camera (EFOSC). We used the MOS mode of
EFOSC to obtain the spectra of 9 of the emitting line knots. The
EFOSC spectrograph was used with a 300 gr/mm grating and the $\rm
2048\times2048$ thinned Loral CCD detector, which provided
coverage of the spectral region $\rm 3860-8070 \AA$. Slit widths
of 1.75" yielded a resolution of $\rm \sim 19\AA$. We obtained
eleven exposures of 1530 sec, for a total of $\rm
\sim 4.65$ hours. \\
The TNG observations were taken in the photometric nights of 26th March
and 22nd April 2004 with DOLORES. We
used the MOS mode of DOLORES to obtain the spectra of 8 of the
emitting line knots and the nuclear region of CGCG97-125 \citep{ZWHE61}.
The DOLORES spectrograph was used with a
grating which provided coverage of the spectral region $\rm 3200-8000\AA$.
Slit widths of 1.6" yielded a resolution of $\rm \sim
17\AA$. We obtained six exposures of 1800 sec, for a total of $\rm 3$ hours.
All the emitting line regions observed in MOS spectroscopy are 
shown in Fig.\ref{ON}\\
In addition we took long slit drift-scan spectra of the bright galaxies CGCG97-114 
and CGCG97-125 using the Loiano/1.52 m and the OHP/1.93m telescopes respectively. 
At Loiano, the BFOSC spectrograph was used with a 300
gr/mm grating and $\rm 1300\times1340$ thinned EEV CCD detector,
which provided a spectral coverage $\rm 3600-8900 \AA$. 
A slit width of 2.00" yielded a resolution of $\rm \sim 20\AA$. 
At OHP, the CARELEC spectrograph was used with a 300
gr/mm grating and $\rm 2048\times1024$ EEV CCD detector,
which provided a spectral coverage $\rm 3400-7000 \AA$.
A slit width of 2.50" yielded a resolution of $\rm \sim 10\AA$. 
Both the observations were taken in the "drift-scan" mode, with the
slit parallel to the galaxy major axis, drifting over the optical
surface of the galaxy. 
The total exposure time was 2400 sec for CGCG97-114 and 1800 for CGCG97-125 respectively.\\
The reduction of all the spectra was carried out using standard tasks
in the IRAF package. 
The $\lambda$ calibration
was carried out using 
exposures of He/Ar lamps for each slit. 
Typical errors on the
dispersion solution are of $\rm \sim 0.5-1 \AA$, as confirmed by
the measurements of the sky lines. 
Spectra were flux-calibrated using the spectrophotometric standard
stars: ltt 3864 for the ESO, Feige 67 for the TNG and
Feige 34 for the Loiano and the OHP observations.\\
The redshift of each knot was derived as the mean of the
individual redshift obtained from each emission line. Our results
are shown in Tab. \ref{vel} and compared
with the previous measurements by \cite{SAKK02} and \cite{GAVC03}.
Due to the low resolution of our MOS (R$\sim$300), the mean 
redshift uncertainty is $\geq 150 \rm km/s$, 
consistent with the discrepancies between the various entries in Tab. \ref{vel}.
Therefore the present observations cannot be used for a detailed dynamical analysis of the group.

\subsubsection{Line measurements}
\begin{figure*}
\centering
\includegraphics[width=15cm]{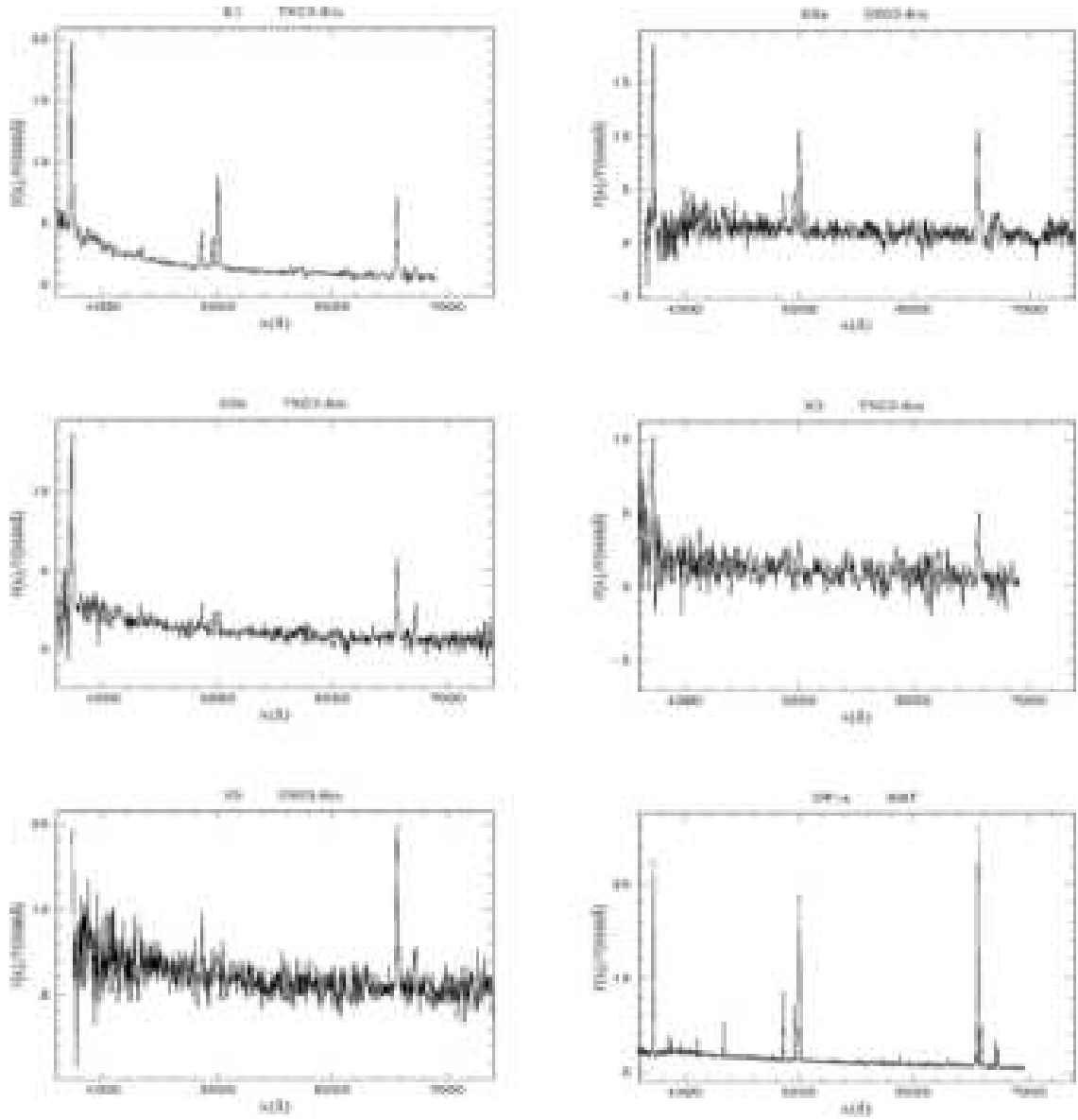}
\small{\caption{The observed, smoothed (step 3), one dimensional
spectra. The object identification and telescope are labeled on
each panel.} \label{spec}}
\end{figure*}

All spectra were shifted to the rest frame wavelength and
normalized to their intensity in the interval 5400-5600 $\rm \AA$.
The flux-calibrated, normalized spectra are presented in Fig.
\ref{spec}. Under visual inspection of the spectra we carried out
the measurement of the emission lines. 
H$\alpha$ ($\lambda 6563$) is bracketed  by the weaker
[NII] doublet ([NII1] $\lambda 6548$ and [NII2] $\lambda 6584$).
The three lines are not well resolved, thus using the task SPLOT
we performed a two Gaussian fit to the blended emissions providing
an estimate of the line ratio $\rm[NII]\lambda6584/(\rm H\alpha +
[NII]\lambda6548)$. The two bright galaxies CGCG97-125 and
CGCG97-114 show evidence for underlying Balmer absorption. 
We de-blended the underlying
absorption from the emission lines as discussed in \cite{gavspectra}.
From H$\beta$ and H$\alpha$ corrected for de-blending from [NII] we evaluate the
Balmer decrement (assuming $T$=10000K and $n$=100 $\rm e/cm^3$):
\begin{equation}
C1 = \frac{log(\frac{1}{2.86} \times \frac{L_{H\alpha}}{L_{H\beta}})}{0.33}
\end{equation}
(in the current notation: A($H\beta$)=2.5*$C_1$).\\
The corrected line fluxes are derived, relative to H$\beta$, using $C_1$  and the de-reddening
law $f(\lambda)$ of \cite{lequex}. 
When H$\beta$ is undetected (only in the DW2b spectrum taken at ESO) we derive a $3*\sigma$
lower limit to $C_1$ using \citep{gavspectra}:
$H\beta<3\times RMS_{(4500-4800)}\times H\alpha HWHM$
assuming that H$\alpha$ and H$\beta$ have similar HWHM (Half
Width Half Maximum).\\
In order to compare our observations with the ones presented by \cite{SAKK02} 
we re-measured, using the method
described above, the 1D spectra obtained from the 
6.5m MMT observations and kindly provided by these authors.
The two sets of measurements presented 
in Tab. \ref{lines} are in fair agreement.

\subsection{High Resolution spectroscopy}
We obtained high dispersion 
long-slit spectra of CGCG97-125 and CGCG97-120 with the $\rm 1.93~m$ telescope of the 
Observatoire de Haute Provence (OHP), 
equipped with the CARELEC spectrograph 
coupled with a $2048 \times 512$ TK CCD, giving a spatial scale of 0.54 arcsec per pixel. 
The observations were carried out in the night of April 20, 2004 in 
approximately 2 arcsec seeing conditions through a 
slit of $5~{\rm arcmin}\times 2$ arcsec. 
The selected grism gives a spectral resolution of 33 $\rm \AA/mm$ or 0.45 $\rm \AA/pix$ 
and R$\sim$5400 in the region 6080-6990 \AA ~containing the redshifted 
$\rm H\alpha$ ( $\rm\lambda~6562.8~\AA$), the [NII] doublet 
($\rm\lambda~6548.1,~6583.4~\AA$) and the [SII] doublet 
($\rm\lambda~6717.0, 6731.3~\AA$).

\section{Results}
\subsection{The H$\alpha$ trails}
\label{plasma}
\begin{figure*}
\centering
\includegraphics[width=19cm]{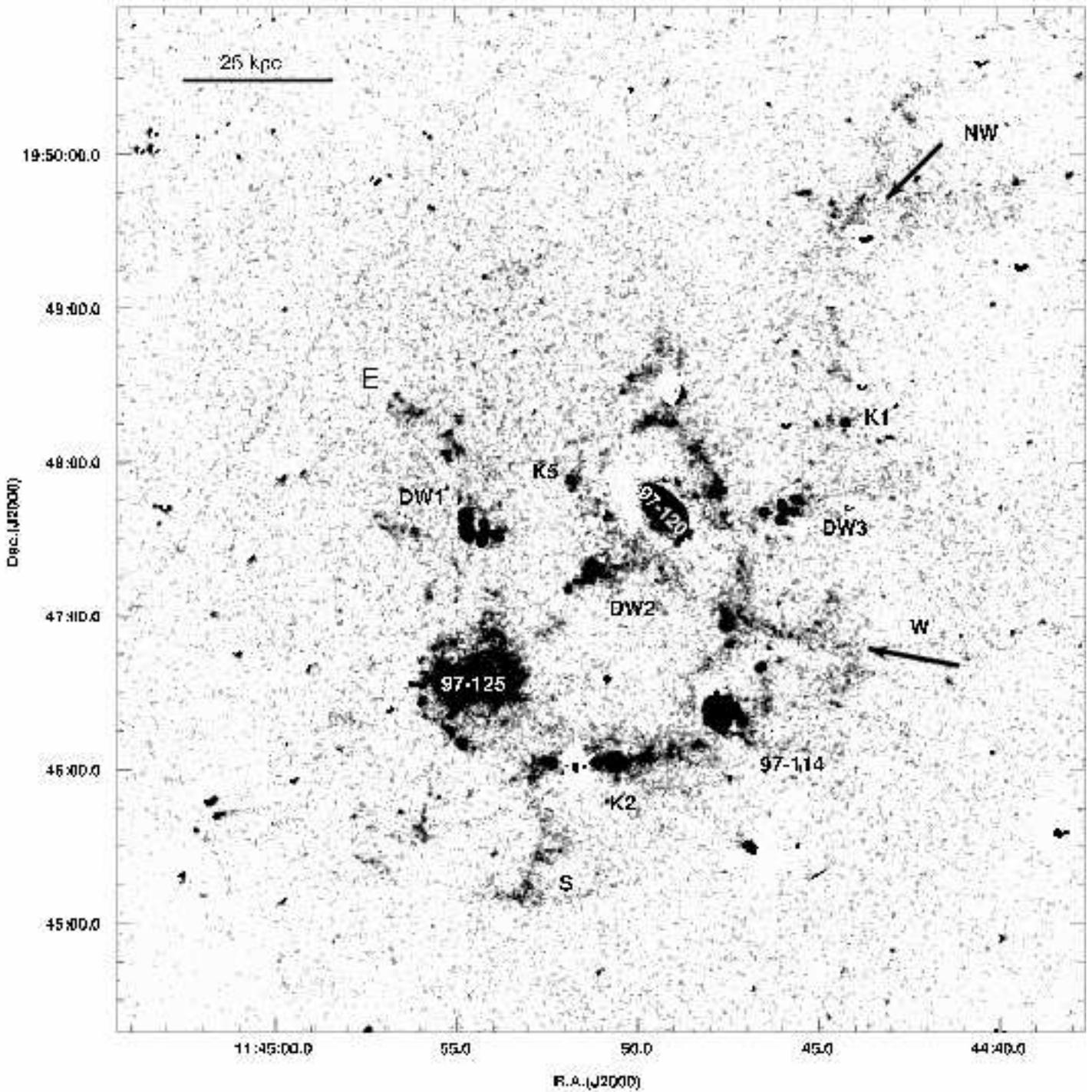}
\small{\caption{Extended low surface brightness loops and trails (labeled NW, W and E)
appear in the H$\alpha$+[NII] NET frame around the giant and dwarf galaxies of BIG.}
\label{net}}
\end{figure*}

\begin{figure*}
\centering
\includegraphics[width=19cm]{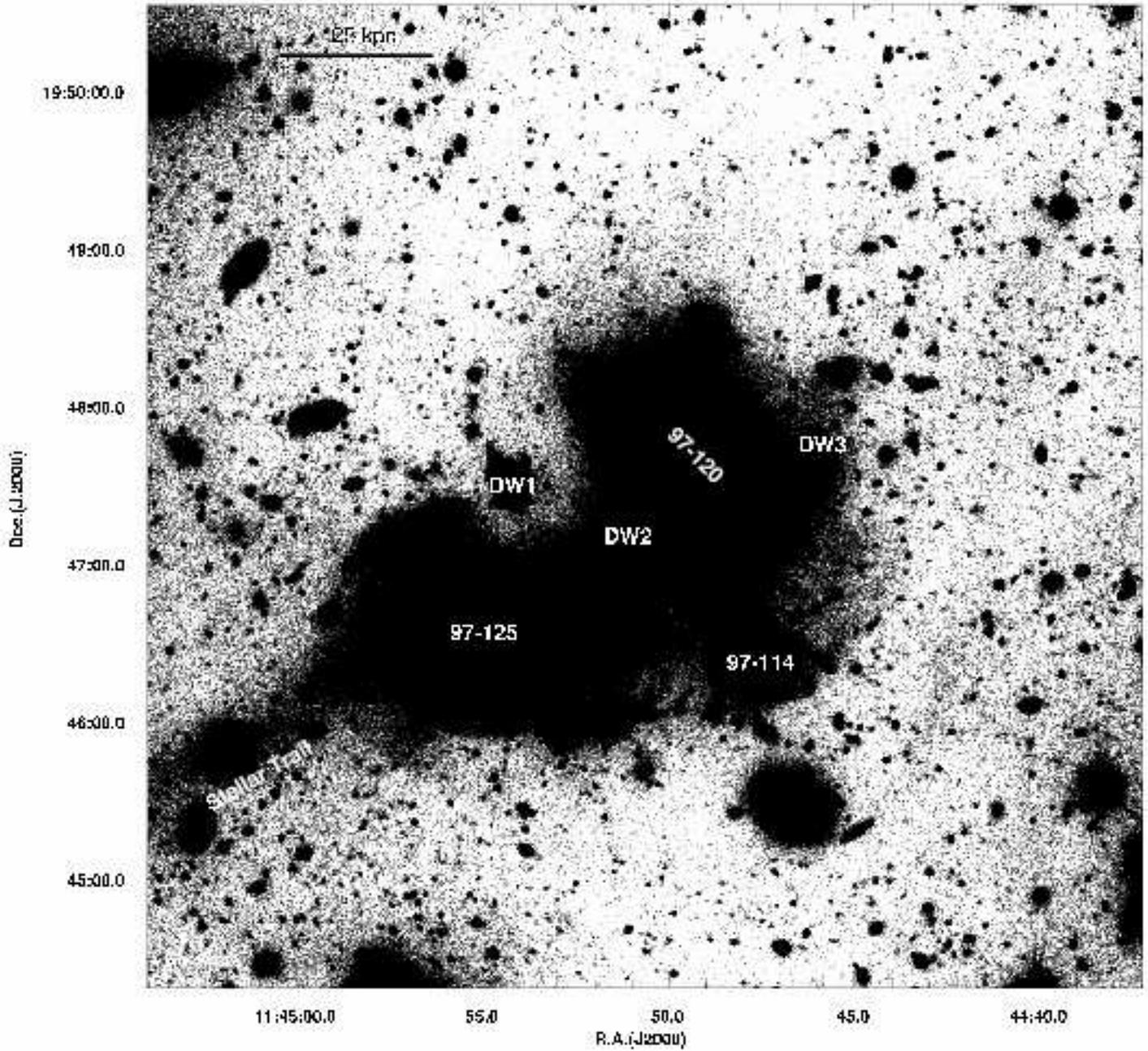}
\small{\caption{A stellar trail starting from CGCG97-125 and diffuse intra-group light appear in the high 
contract $r'$ band image of the BIG group.}
\label{Rcontrast}}
\end{figure*}

The new H$\alpha$ images of BIG obtained at the TNG 
reveal for the first time a spectacular 
H$\alpha$ filamentary structure on top of which 
the star forming knots previously observed by \cite{SAKK02} and by \cite{GAVC03} 
represent the tip of the iceberg (see Fig. \ref{net}). 
Multiple loops of ionized gas appear with a projected length 
exceeding 150 kpc and a typical transverse size of 5 kpc, making these among
the most extended low-brightness H$\alpha$ emission features ever detected.
One stream (labeled NW in Fig. \ref{net}) extends from the northern edge
of the frame to the dwarf galaxy DW3, with an extension of $\sim100 \rm kpc$. 
The second and brightest one (labeled W in Fig. \ref{net}) apparently traces a loop around the galaxy
CGCG97-120 and seems connected to the bridge (labeled K2 in Fig.\ref{net})
between CGCG97-114 and CGCG97-125. 
If this is the case, the total projected extension of the NW and W trails would result
$\rm \sim150~kpc$.
The looped trajectory traced by the trails in Fig.\ref{net} should imagined as stretched
along the line of sight, similar to a ``bottle opener'' that 
travels through a cork at a small angle $\theta$ with the line of sight.
In this case, the projected length of the trails would be amplified by $tan^{-1}$ $\theta$.
In addition to the filamentary features, at least two other
diffuse H$\alpha$ regions (labeled S and E in Fig.\ref{net})
are detected.\\ 
The total diffuse ($\rm
H\alpha+[NII]$) emission (e.g. excluding the contribution of the
three bright galaxies and of the ten dwarfs/HII regions previously
discovered) results $\sim 1.2\times10^{-13} \rm ~erg~ cm^{2} ~s^{-1}$,
similar to the flux collected from CGCG97-125, and the typical
surface brightness is $\rm 10^{-17.6}-10^{-18.3} ~erg ~cm^{-2} ~s^{-1} ~arcsec^{-2}$.\\
The loop around CGCG97-120 alone contributes 
$\sim 2.4\times10^{-14} \rm ~erg~ cm^{2} ~s^{-1}$,
obtained integrating the $\rm H\alpha+[NII]$ emission in a circular corona of 10 kpc radius and
an annulus of 5 kpc centered on CGCG97-120.
The derived line intensity is 2.05 Rayleigh (1 Rayleigh = $\rm 10^{6}/4\pi ~photons ~cm^{-2} ~s^{-1} ~sr^{-1}$),
corresponding to an emission measure (EM) of 5.7 $\rm cm^{-6} ~pc$.
Assuming a torus geometry with a circular section of radius $\rm \sim5~kpc$ and a filling factor of 1, 
the resultant plasma density is $n_{e}\sim 3.3\times10^{-2} \rm cm^{-3}$ and the ionized column density is
$N_{e}\sim 5 \times 10^{20}~\rm cm^{-2}$ (the inferred densities would be higher if the gas is in clumps or filaments, which
is likely).
The emission measure in the NW trail is $\sim 1.3 \rm ~cm^{-6}$ pc 
with a plasma density of $n_{e} \sim 1.1 \rm ~cm^{-3}$.\\
The trails' geometry is strongly suggestive of a Rosetta orbit, typically found with the tidal disruption 
of a satellite galaxy. 
Contrary to other known examples of tidal streams 
(i.e. \citealp{ibata01,shang98,wehner05}) the features observed herein show 
strong H$\alpha$ emission, but it is difficult to determine whether or not the features are associated with a diffuse 
stellar component.
The loop around CGCG97-120 lies in fact within the galaxy's optical radius, 
making impossible any estimate of the continuum emission associated with the H$\alpha$ feature. 
A diffuse intra-group light, with typical surface brightness 
$\mu_{r'}$ between 25.2 and  26.6 $\rm mag ~arcsec^{-2}$, is detected in the center 
of the group (but not around DW1) and in
its western part (where the W H$\alpha$ trail starts), while no continuum emission 
above $\mu_{r'}\sim\rm 27.4~ mag ~arcsec^{-2}$ is associated with the NW and S H$\alpha$ trails (Fig.\ref{Rcontrast}), 
implying an extremely high H$\alpha$ equivalent width (E.W.($\rm H\alpha+[NII]$) $\geq 100-150 ~\rm \AA$).
It is also interesting to note that the one clear example of tidal 
stellar trail ($\mu_{r'}\sim\rm 26~ mag ~arcsec^{-2}$) 
detected in this region and extending from CGCG97-125 to south-east (see Fig.\ref{Rcontrast}) does not 
show H$\alpha$ emission.\\
Is there neutral gas associated with the H$\alpha$ trails?
The new HI observations obtained at Arecibo, combined with the higher resolution HI map
previously available from \cite{SAKK02}, give additional hints 
on the properties of these extraordinary features.
Fig.\ref{HI2} shows two HI spectra: one (solid line) obtained averaging three (non-independent)
spectra from regions 97-125, 97-120, 97-114 (Fig.\ref{HIpoints}), the other (dotted line) 
obtained from the independent pointing BIG-NW. 
In the combined spectrum we detect $\sim 6.5 \times 10^{9}\rm M_{\odot}$ in the two-horn peak
at $8000<V<9000~\rm km~s^{-1}$, consistent with $\sim 3.9 \times 10^{9}\rm M_{\odot}$ from
CGCG97-125,  $\sim 3 \times 10^{8}\rm M_{\odot}$ from CGCG97-114 and $\sim 2.2 \times 10^{9}\rm M_{\odot}$ from
a diffuse component, possibly associated with DW2 and K2, as determined by \cite{SAKK02}. 
However, in the same combined profile we detect a new component at 
$7500<V<8000~\rm km~s^{-1}$ with an HI mass $\sim 1 \times 10^{9}\rm M_{\odot}$.
In the same velocity interval, another $\sim 7 \times 10^{8}\rm M_{\odot}$ 
($N_{HI}\sim 3 \times 10^{20}~\rm cm^{-2}$) component
centered at $\sim 7870 \rm~ km~s^{-1}$ is detected in the independent BIG-NW region.
In the velocity range $7500<V<8000~\rm km~s^{-1}$ there are no galaxies 
(see Table \ref{vel} and Fig. \ref{ON}) in the observed beams
(nor within their side-lobes) that could possibly contribute to this HI detection. 
The additional fact that this component, characterized by a width 
$ \Delta V \sim 100 \rm km~s^{-1}$, in spite of being included in the data-cube
of \cite{SAKK02}, was not detected above their limiting sensitivity of 0.4 mJy/beam 
implies that it extends over an area $\geq 1~ \rm arcmin^2$.
We argue that the new low-velocity HI component is associated with the diffused H$\alpha$ trails, 
suggesting that the gaseous trails of BIG contain both neutral and ionized phases.

\subsection{The extragalactic star forming systems}

\subsubsection{Current star formation}
Contrary to the gaseous filaments, current star formation is clearly 
observed in all compact HII regions and dwarf and giant galaxies comprising BIG, 
suggesting that bursts of star formation are presently taking place within this group.
At least ten star-forming regions are associated with 
dwarf systems (or extragalactic HII regions) with E.W.($\rm H\alpha+[NII]$) 
often exceeding 100 \AA~(see Table \ref{tabha}).
These values are typical of HII regions \citep{kenn89,bresolin} and usually observed 
in tidal dwarf candidates in interacting systems \citep{duc98,mendes04}.
However, as remarked by \cite{SAKK02}, 
it is the first time that such a high density of star-forming 
galaxies has been seen in a nearby cluster, 
in spite of their having collected data over an area of A1367, 
Coma, and the Virgo cluster approximately 500 times larger than the group size 
\citep{catinella,gavha,haanna,ha06,jorge,bosha2,bosgav02}.\\
It is interesting to note that the dwarf galaxy DW2, and in particular the knot DW2c,
shows clear Post-Starburst signatures in its spectrum, with low residual current star-formation, 
extremely blue continuum (B-R $\sim0.16\pm0.20$ mag), strong Balmer 
absorption (EW(H$\delta$)$\rm \sim 8 \AA$) and $\rm[OII]$ and H$\alpha$ in emission (see Fig.\ref{spec}).
These features suggest that the bulk of the starburst took place already $\leq 10^{8}$ years ago 
(e.g. \citealp{poggia97,poggia99,kauff03}).

\subsubsection{Kinematics}
\label{seckin}
\begin{figure*}
\centering
\includegraphics[width=8cm]{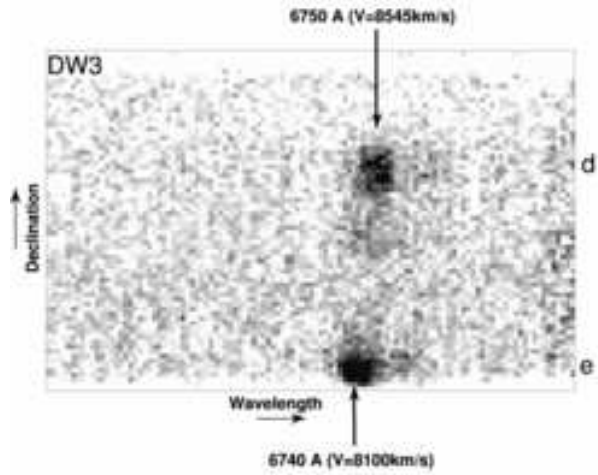}
\small{\caption{The low resolution 2D spectrum obtained at ESO/3.6 for knots DW3d and DW3e, shows 
a significant difference ($\sim500 \rm km~s^{-1}$) in the their velocity.}
\label{dw3rot}}
\end{figure*}

Table \ref{vel} lists the positions and radial
velocities of the objects measured spectroscopically (see also Fig.\ref{ON}).
Our observations confirm the physical association of all the emitting line 
objects within the bright galaxies CGCG97-114 and CGCG97-125 \citep{ZWHE61}.
The velocity of galaxies in BIG ($<V>=8230~ \rm km~s^{-1}$ and $\sigma_V=170~
\rm km~s^{-1}$) significantly exceeds the mean cluster velocity  
($<V>=6484~ \rm km~s^{-1}$, $\sigma_V=891~\rm km~s^{-1}$, \citealp{COGA04}), 
suggesting infall at $\sim$ 1700 $\rm km~s^{-1}$ into the cluster core.\\
Although the MOS observations cannot be used for a detailed 
dynamical analysis of the group's members (due to their low resolution), they provide 
some information regarding the internal dynamic of DW3.
DW3, in fact, presents remarkable kinematic peculiarities, as shown in Fig.\ref{dw3rot}, 
where the two emitting line knots DW3-d and DW3-e happened to lie within the same slit of 
one of our MOS observations taken at the ESO/3.6m telescope.
A relative velocity of $\sim 450\rm ~km~s^{-1}$, unaffected by the uncertainty in the 
wavelength calibration\footnote{We cannot however exclude that the different position of the two knots 
on the slit contributes to the observed velocity offset.}
 is observed between the two knots. 
This high velocity gradient suggests that the two knots do not form a gravitationally bound virialized system.

\subsubsection{Metal abundances}
\begin{figure*}
\centering
\includegraphics[width=9.5cm]{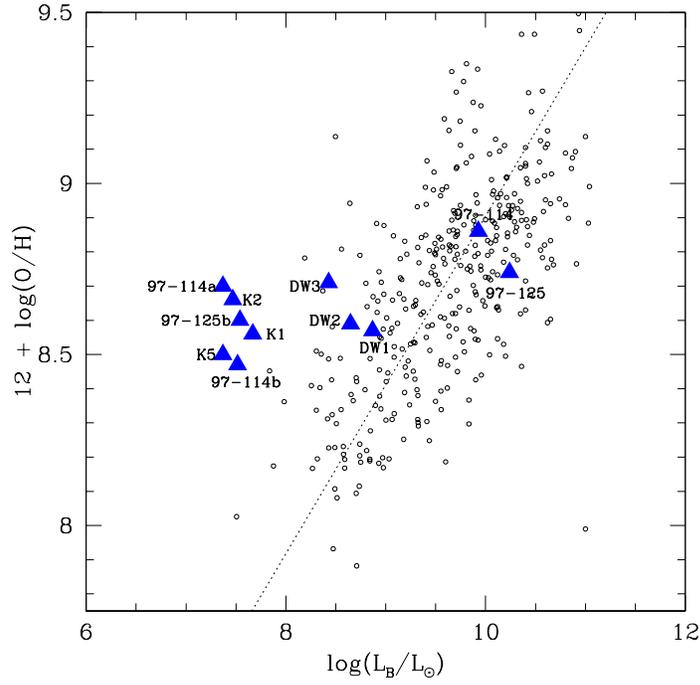}
\small{\caption{The relation between metallicity and B-band luminosity (with linear best-fit)
for galaxies in nearby clusters
(empty circles, adapted from Gavazzi et al. 2004). 
The triangles mark the mean metallicity obtained for the
star forming systems of BIG.}
\label{l_mrel}}
\end{figure*}

In order to determine the metal content of BIG's constituents,
we average five different empirical determinations based on the following line ratios:
$\rm R_{23} \equiv ([OII]\lambda3727+[OIII]\lambda4959,5007)/H\beta$
\citep{zaritsky94,mcg91}, $\rm [NII]\lambda6583/[OII]\lambda3727$
\citep{kewley02}, $\rm [NII]\lambda6583/H\alpha$ \citep{vanzee98} and
$\rm [OIII]\lambda5007/ [NII]\lambda6583$ \citep{dutil99}.
Since the $\rm R_{23}$ is not a monotonic function of the oxygen abundance, 
we used the line ratio $\rm [OIII]\lambda5007/ [NII]\lambda6583$ to distinguish between the 
lower and upper branches of the calibration, as suggest by \cite{kobulnicky99}.
The metallicities obtained from the various methods are shown individually and averaged in Table~\ref{metals}.\\
All the star-forming regions in BIG are metal-rich, with$\rm 8.5< 12+log(O/H) < 8.9$, and therefore
extend the findings of \cite{SAKK02}.\\
It is well known that irregular and spiral galaxies follow a ''metallicity - luminosity
relation'' \citep{skillman89}.
Fig.\ref{l_mrel} shows this relation for galaxies in the Virgo cluster 
(circles, obtained by Gavazzi et al. in prep. 
using the same methods and calibrations adopted here) and
for the star-forming systems in BIG (triangles).
The two brightest galaxies in the group (CGCG97-114 and CGCG97-125) 
have a normal metal content for their luminosity.
Conversely the faintest ($L_{B}< 10^{8} \rm L_{\odot}$) star-forming knots show higher abundances 
than expected from their intrinsic luminosities.
If these systems (K1, K5, K2, 97-114a, 97-114b and 97-125b) were isolated, independently-evolved,
dwarf galaxies, we would measure a metallicity 0.6-1.2 dex lower than observed.
Moreover their abundances are consistent with the values measured for tidal dwarf
systems ($\rm 12+log(O/H) \sim 8.60$, independent from their absolute magnitude
-- e.g. \citealp{duc99,duc00}).
At intermediate luminosities, the dwarf systems DW1, DW2 and DW3 have a high-metal 
content but are still consistent, within the calibration uncertainties,
with the abundances observed in typical dwarf galaxies of their luminosity.
However (as discussed in the Section \ref{seckin}) we are not sure whether the star forming region
DW3 should be considered as an individual galaxy or as an unbound collection of HII regions. 
In the latter case their metallicity would result  $\sim 0.8$ dex higher than expected from the 
metallicity-luminosity relation.\\
As a result of the above considerations, we conclude that the HII regions in BIG are normal, independently evolved dwarf galaxies, 
reinforcing the scenario proposed by \cite{SAKK02} that these systems formed from enriched
material stripped by tidal interactions from the two brightest galaxies in BIG.

\subsection{The brightest group member: CGCG97-125}
\begin{figure*}
\centering
\includegraphics[height=6cm]{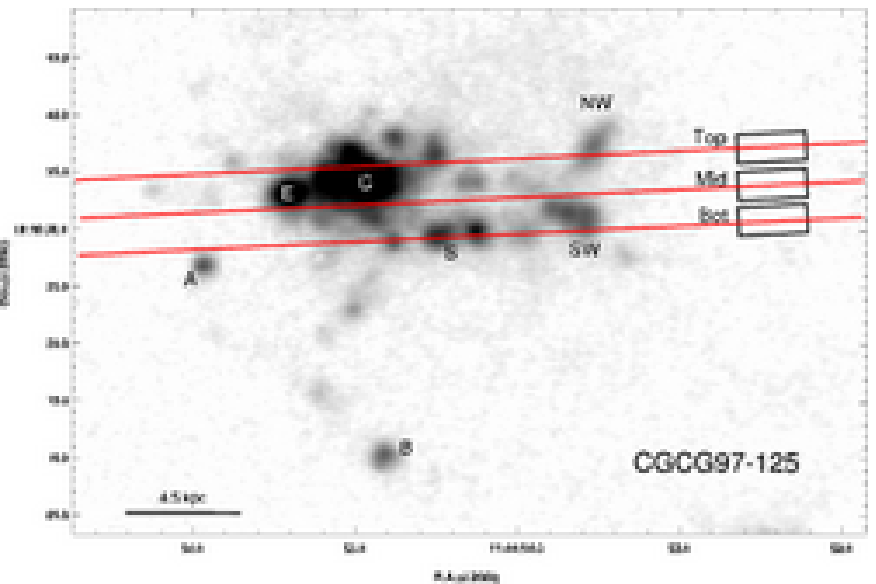}
\includegraphics[width=7.5cm, height=6.5cm]{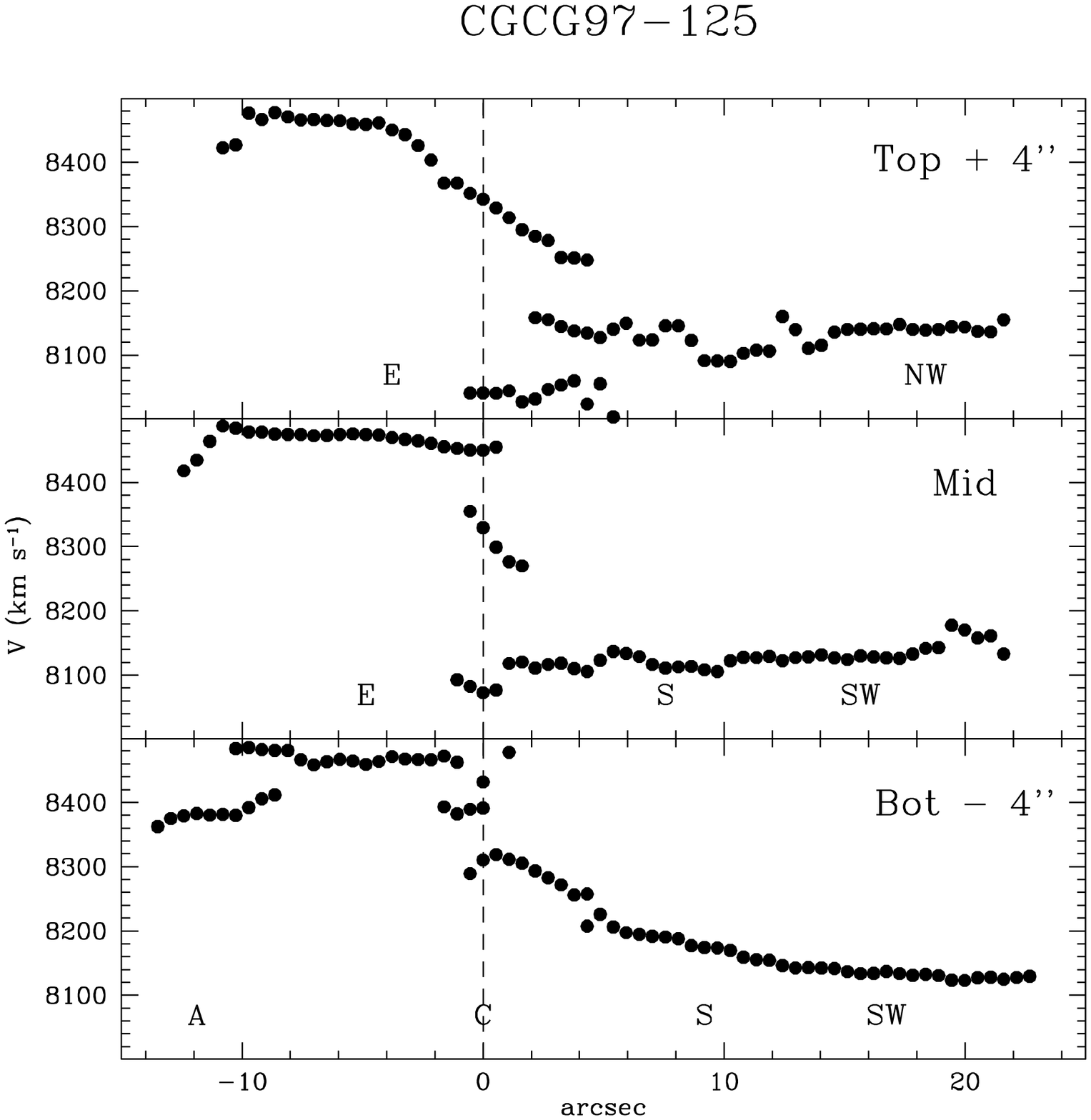}
\small{\caption{ Left panel: The position and the width (rectangular areas on the right) of the three 
slits obtained for CGCG97-125. The slits are superposed to the $\rm H\alpha+[NII]$ net image.
Right Panel: The  rotations curves obtained for the three indicated positions of CGCG97-125.}
\label{rotslit}}
\end{figure*}

\begin{figure*}
\centering
\includegraphics[width=9.5cm, height=7cm]{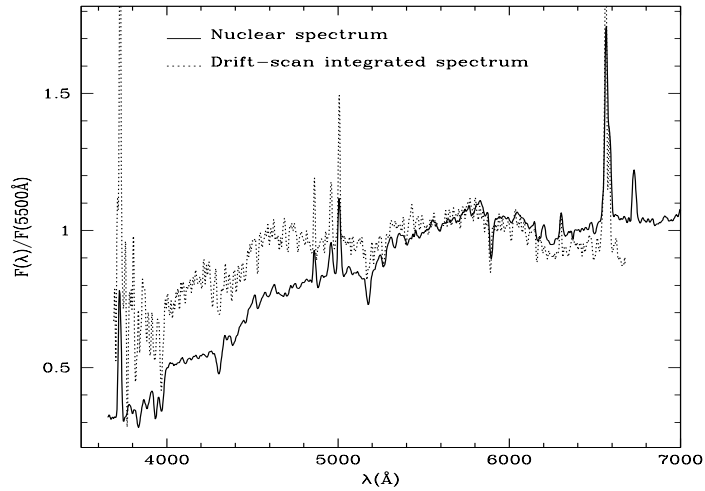}
\small{\caption{Comparison between the drift-scan integrated (dotted) and nuclear (solid) spectrum 
of CGCG97-125.}
\label{spec125}}
\end{figure*}

\begin{figure*}
\centering
\includegraphics[width=12.5cm]{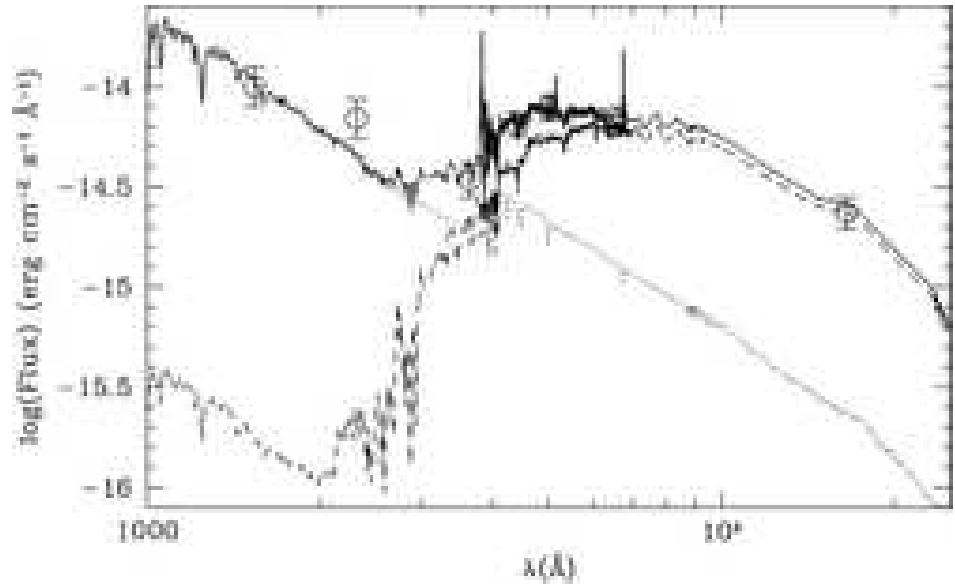}
\small{\caption{The SED of CGCG97-125, corrected for internal extinction. 
Solid thick lines indicate the optical nuclear and drift-scan spectra. 
Circles indicate the photometric observations and their uncertainties. 
Best fitting models for the nuclear spectrum (dashed line) and for the starburst component (dotted line) 
are given. The solid thin line shows the resulting best fitting SED for CGCG97-125.}
\label{sedfinale}}
\end{figure*}

CGCG97-125 is classified as S0a \citep{goldmine}, 
consistent with its red B-R color index ($\sim$ 1.34 mag) and with the 
shape of the continuum optical spectrum (Fig.\ref{spec125}).
However, this system is far from being a normal early type galaxy.
The presence of stellar shells around CGCG97-125 (see Fig.\ref{R})
clearly indicates a past interaction/merger event. 
Numerical simulations predict that the stars from a satellite make a system
of shells several $10^{8}$ yr after the end of the merging event
which lasts for more than 1 Gyr \citep{shell}.\\
This scenario is reinforced by the unusual dynamical state 
of the galaxy, as revealed by the high resolution spectra obtained at the OHP telescope.
Velocity plots of CGCG97-125 were extracted from each spectrum 
by measuring the wavelength of the H$\alpha$ line in each pixel 
along the slits.  
The three rotation curves so obtained are given in Fig.\ref{rotslit}. 
In each diagram the recessional velocity is plotted as a function of position 
along the slit (the spatial axis runs from E (left) to W (right)). 
All three spectra show multiple velocity components near the galaxy center (C) 
where sudden velocity jumps of $\sim$100-150$\rm km~s^{-1}$ are detected. 
Even though examples of kinematic disturbances have been observed in normal galaxies \citep{rubin99,haynes00}
 and interacting systems \citep{jore,duc98}, features similar to those 
observed in CGCG97-125 were found only in UGC6697 \citep{ugc6697}, a merging 
systems in the NW region of Abell1367.\\ 
Moreover, CGCG97-125 shows strong emission lines (Fig.\ref{spec125}) 
which are never seen in galaxies earlier than Sc (see e.g.\citealp{gav02}),
indicating that this galaxy is still experiencing a strong burst
of star formation, likely produced by the merging event -- a kind of \emph{rejuvenated} early type galaxy.
If we compare the continua in the drift-scan integrated spectrum and the nuclear 
spectrum of CGCG97-125 we find that 
the first appears considerably bluer than the latter (see Fig.\ref{spec125}).
Assuming that the nuclear continuum is representative of the old stellar population
in this galaxy, pre-existing the burst, and that the drift-scan continuum is affected by the burst, 
we argue that the difference between the continua traces the recent star formation history of this galaxy, which
can be therefore estimated.
We use the SED fitting procedure proposed by \cite{gav02} and further developed by \cite{paolotesi}.
We assume a star formation history (SFH) "a la Sandage":
\begin{equation} 
SFH(t,\tau)=\frac{t}{\tau^{2}}\times\exp(-\frac{t^{2}}{2\tau^{2}})  
\end{equation}
and the \cite{BC2003} (BC03) population synthesis models.
We fit the nuclear spectrum of CGCG97-125, corrected for extinction\footnote{
We assume that the nuclear and integrated spectra are affected by the same amount of attenuation, as 
supported by the similar value of $C1$ obtained in the two spectra (see Table \ref{lines}).
We thus estimate A($\lambda$) for the integrated SED using the ultraviolet spectral slope $\beta$ as discussed 
in \cite{COdust05} and assuming the Milky-Way attenuation curve \citep{bosellised}.}, 
assuming t=13 Gyr, a Salpeter IMF ($\alpha$ = 2.35 from 0.1 to 100 $\rm M_{\odot}$; \citealp{salpeter}),
and exploring a parameter grid in metallicity (Z), 
from 1/50 to 2.5 $\rm Z_{\odot}$ in five steps: 0.0004, 0.004, 0.008, 0.02, and 0.05.  We also let  
$\tau$ to vary from 0.1 to 25 Gyr in 45 approximately logarithmic steps.
The best fit parameters obtained using the BC03 models are summarized in Table \ref{fitnucleo}.
The best value of $\tau$ is consistent with the one ($\tau\leq3.1$ Gyr) obtained by \cite{gav02} 
fitting a template of S0 galaxies. This result confirms that the continuum of 
the nuclear spectrum is dominated by an old stellar population of the same age expected for an unperturbed S0.
By subtracting the best fit nuclear SED from the integrated SED of CGCG97-125\footnote{Broad band 
integrated photometry of CGCG97-125 is taken from \cite{CORBGA05} and \cite{goldmine}.}, we can estimate the 
starburst contribution to the galaxy emission, the burst age and mass.
The best-fitting parameters obtained for the burst SED are summarized in Table \ref{fitnucleo}.
The burst age results $\sim$1.4-2.8 Gyr and the stellar mass produced during the burst is 
$\sim2\times10^{9}~\rm M_{\odot}$, 
consistent with the values previously obtained from independent estimates.
The resulting best fit SED for CGCG97-125 is shown in Fig.\ref{sedfinale} (solid line).
We thus conclude that the star burst in CGCG97-125 was: initiated
$\sim$1.5 Gyr ago, probably produced by the minor merging of a $\sim2\times10^{9}~\rm M_{\odot}$ satellite, 
and is still taking place.
Our result points out that a minor merging, while producing only a small 
fraction ($\sim$2\%) of new stars, is
sufficient to disturb the morphology and the dynamics of a giant galaxy.

\section{Discussion}
\subsection{The origin of the H$\alpha$ trails}
\label{origtrail}
\begin{figure*}
\centering
\includegraphics[width=7cm, height=5cm]{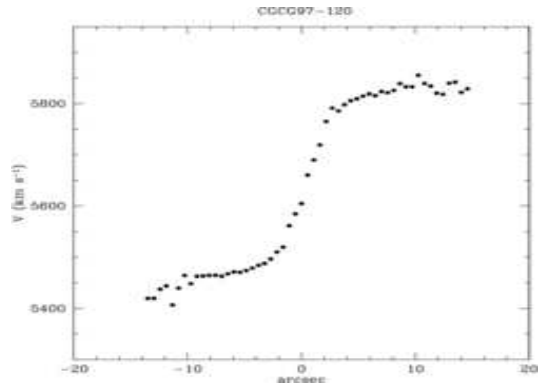}
\small{\caption{The H$\alpha$ rotation curve of CGCG97-120.}
\label{rot120}}
\end{figure*}

The high star formation associated with the galaxies in BIG, the high 
metal content of the extragalactic HII regions, and the unusual properties of CGCG97-125, are
observed in typical compact groups and are strongly suggestive of tidal interaction between 
the group's members. Therefore it is certain that gravitational interactions are strongly 
influencing the evolutionary history of this group.\\
Are the multiple trails of ionized gas consistent with a tidal interaction scenario?
One of the H$\alpha$ trails apparently traces a perfect loop around CGCG97-120 (see Fig.\ref{net}). 
Yet this galaxy seems not to be associated with 
BIG, having a recessional velocity of 5635 $\rm km~s^{-1}$, thus blue-shifted 
with respect to A1367 by approximately 800 $\rm km~s^{-1}$. Is this a pure coincidence?
The galaxy's morphology and kinematics are completely 
unperturbed, showing no signs of interaction (see Fig.\ref{R} and Fig.\ref{rot120}). 
Moreover the scattering angle between a satellite galaxy and CGCG97-120 would 
be only $\sim$4-10 degrees, assuming a classical scattering model and an impact 
parameter of $\sim$10 kpc, too small to produce the observed loop.
Thus we must assume that the association between 
CGCG97-120 and the H$\alpha$ trails is a mere coincidence.\\
Whatever the role of CGCG97-120 is, it is indisputable that the geometry of the 
H$\alpha$ trails is strongly suggestive of a Rosetta orbit typical of the tidal disruption 
of a satellite galaxy. 
Contrary to other known examples of tidal streams 
(i.e. \citealp{ibata01,shang98,wehner05}), 
the features here observed show strong H$\alpha$ emission. 
Tidal streams discovered in interacting systems (e.g. \citealp{shang98,forbes03}) are in fact detected only 
in continuum with no H$\alpha$ emission, even if associated to strong starburst merging systems \citep{wehner05}.
The case offered by BIG seems therefore unique among interacting systems 
as it is characterized by strong emission line trails.\\
Even if they have never been observed among other interacting systems, similar H$\alpha$ trails have been recently discovered 
in the NW part of Abell1367 itself. \cite{GAVB01} discovered two low surface brightness 
H$\alpha$ cometary tails, with a projected length of 75 kpc, associated with two star forming 
systems: CGCG97-073 and CGCG97-079. 
The morphology of these tails (whose typical size and gas densities are similar 
to the trails observed in BIG) suggests that galaxies in the NW group are experiencing ram pressure 
due to their high velocity motion through the IGM \citep{GAVB01}. 
The difference between the case of CGCG97-079 and CGCG97-073 and BIG is that the former are isolated 
systems infalling into the cluster, while galaxies in BIG are infalling within a compact 
group, where gravitational interactions are not negligible. 
For these reasons, the unique features observed in BIG lead to the supposition that it is not only tidal interaction 
which can produce the H$\alpha$ trails but that probably the mutual influence 
of tidal and ram pressure can also explain 
the properties of the H$\alpha$ trails, unless we are dealing with a unique
example of tidal trail with associated ionized gas,
on scales exceeding 150 kpc.
Since BIG is infalling at 1700 $\rm km~s^{-1}$ into the center of Abell1367, 
ram pressure is likely playing a non-negligible role on the evolution of this group.\\    
This scenario seems supported by the new HI observations obtained at Arecibo.
The HI profile indicates the presence of diffused neutral gas probably associated 
with the H$\alpha$ trail. This diffuse component show a recessional velocity in the range between 7500 
and 8000 $\rm km~s^{-1}$ (systematically lower than the typical velocity of the star forming 
systems), as if the gas was ram-pressure stripped from galaxies infalling at 1700 $\rm km~s^{-1}$ and experienced 
a progressive velocity decrease
due to the friction with the cluster IGM (V$\sim 6500~ \rm km~s^{-1}$).
In this perspective the properties of the  HI diffused component in BIG are not too different from
the ones observed in the extended 
HI plume recently discovered in Virgo and interpreted by \cite{osterloo05} as due 
to ram pressure stripping
from NGC4388.
The H$\alpha$ trails might therefore be suggesting that both mechanisms 
(i.e. tidal interactions among the group members and ram pressure 
with the cluster IGM) are taking place at once, 
as already observed in the Virgo cluster \citep{vollmer4654,vollmer05,vollmer4254}.
Deeper observations and numerical simulations are however 
mandatory to confirm the role of ram pressure stripping on shaping 
the properties of this group.

\subsection{The ionization source of the trails.}
The large extent of the H$\alpha$ trails 
and their associated neutral component indicates that stripped gas can survive from 
100 Myr up to 1.5 Gyr (see Section \ref{plasma}).
This represents a problem for the diffused HI in BIG as well as in other known
extended HI components in clusters (i.e.\citealp{osterloo05}) because HI evaporation by the hot IGM
should occur in few $10^{7}$ yr \citep{COMK77}.
One possibility is that HI evaporation is in fact less effective 
because the gas is shielded by tangled magnetic fields \citep{COMC81,vollmer01}.\\
Furthermore the plasma density derived in section \ref{plasma} 
implies an exceedingly short recombination time for the ionized component along the trails of BIG:
$\tau_{r} = 1/N_{e}\alpha_{a}\sim$ 2-7 Myr, where $\alpha_{a} = 4.2 \times 10^{-13} \rm ~cm^{3} ~s^{-1}$
\citep{osterb89}.
Can this exceedingly short recombination time
of few Myr be reconciled with an age of the phenomenon between some 100 Myr and 1.5 Gyr?
We need a mechanism to keep the gas ionized for such a long time.
In spiral galaxies, if the HI column density is above a few times
$10^{20}~\rm cm^{-2}$, star formation almost invariably occurs \citep{boissier03}.  The
mean column density in the trail is consistent with this value, 
suggesting that the star formation threshold
might locally be exceeded. Hence, star formation could in principle provide the source of
ionization in the trails.
However we have no evidence of star formation along the trails 
and the dynamical picture of BIG is consistent with the idea that at least part 
of the gas has been stripped ionized from the galaxies.\\
A source of ionization for the trails could be provided by UV photons escaping from the star-bursting regions 
within the main group's galaxies, as observed in other 
starburst systems (i.e.\citealp{ngc5252,m82cap,ngc7213}).
However the UV continuum emitted by the two bright galaxies CGCG97125 and CGCG97-114 
could only explain the presence of diffuse ionized gas observed in their surroundings 
($<10-20$ kpc from the center), but it is unlikely 
responsible for the ionization of the H$\alpha$ trails (see also \citealp{veilleux03}).\\
Another possibility is that the gas is ionized by the propagation of shock waves \citep{jog92,barnes04} 
produced during the galaxies collisions and/or the high velocity infall into the hot intracluster medium.
Deep spectroscopic observations are necessary to test this possibility. 
Finally, we must consider the effect of the bremsstrahlung radiation by the ICM on the stripped gas.
Following \cite{vollmer01} and \cite{malon96}
the X-ray ionizing photon flux ($\phi_{i}$) is:
\begin{displaymath}
    \phi_{i} = \frac{ln(0.1/0.0136)}{1.6\times 10^{-9}}
    \frac{F_{X}}{1.5} = 8.3 \times 10^{8}~
    F_{X}~\rm photons~cm^{-2}~s^{-1}
\end{displaymath}
where $F_{X}$ is the X-ray flux. 
Assuming a total cluster X-ray luminosity of $4\times10^{43} \rm erg~cm^{-2}$ \citep{DON98}
  the X-ray flux at a projected distance of $\sim$ 125 kpc from the X-ray center (where BIG is observed)
  is $\sim 2.5\times10^{-6} \rm erg~cm^{-2} s^{-1}$ and $\phi_{i}$ results $\sim 2.1\times 10^{3} \rm
  photons~cm^{-2}~s^{-1}$. 
In equilibrium this gives rise to an ionized column density
$N_{e} = \phi_{i}/\alpha_{a} n_{e}$.
Using $n_{e} = 10^{-2} \rm cm^{-3}$ we obtain $N_{e} \sim 5 \times 10^{20} \rm  cm^{-2}$, 
consistent with value measured in the ionized tails (see Section \ref{plasma}).
This simple calculation shows that the stripped gas
can be kept ionized by the X-ray photons.
A deep spectrum of the trails will help us solving this riddle.

\subsection{The Blue Infalling Group: an unique example among 
compact groups?}

The features observed in BIG are indicative that tidal forces and ram pressure 
are mutually shaping the evolution of this system.
In fact BIG appears as an unique example of a compact group infalling 
into the core of a cluster of galaxies.\\
Is BIG really so different from "normal compact groups"?\\ 
In order to answer to this question, let us compare BIG with 
the prototype of compact groups: the Stephan's Quintet (HCG92, hereafter SQ). This 
group is composed of four interacting galaxies and, lying at approximately the same distance 
of BIG ($\sim$ 92 $h_{70}^{-1}$ Mpc), is ideal for a comparative study.
Both groups show a high number of bright, metal rich 
tidal dwarfs/extragalactic HII regions \citep{IGLV01,mendes01,mendes04} and 
diffuse HI components (with a total HI mass of $M\sim10^{9}-10^{10}$ M$_{\odot}$), suggesting that tidal 
interactions are perturbing the groups' members.
While in the SQ we observe two 
tidal tails with a mean surface brightness of $\mu_{V}=24.4$ and 26 mag $\rm arcsec^{-2}$ \citep{sulentic01} 
respectively, a stellar trail with a typical 
surface brightness of $\mu_{r'}\sim$26 mag $\rm arcsec^{-2}$ and diffuse intra-group light  
(25.2$\leq\mu_{r'}\leq$26.6 mag $\rm arcsec^{-2}$) are observed in BIG.
However BIG shows multiple H$\alpha$ trails and streams with 
a total extension of 150 kpc, features not observed in the SQ \citep{xu99,sulentic01}.
The neutral hydrogen component associated with these streams has a recessional velocity 
in the range between 7500 and 8000 $\rm km~s^{-1}$, and is systematically blue-shifted with respect to the mean 
velocity of the star forming systems.
On the contrary, in the SQ the diffuse HI lies at the same velocity of the star forming regions \citep{wiliamHI}.
As discussed in Sec. \ref{origtrail} the properties of the diffuse H$\alpha$ emission and its associated HI component, 
not observed in the SQ, are consistent with a ram pressure stripping scenario, 
supporting the idea that the differences between BIG and SQ are due 
to the interaction of BIG with the cluster IGM.\\ 
Finally, a galaxy with recessional velocity 
significantly different from the mean group's velocity is observed in both BIG and SQ.
In BIG, CGCG97-120 has a recessional velocity $\sim2500 \rm km~s^{-1}$ lower than the group's 
velocity, while in the SQ NGC7318b is blue-shifted by $\sim900 \rm km~s^{-1}$ with respect to other members.
Multiwavelength observations of SQ \citep{sulentic01,xu99,xu03,xu05} have shown that NGC7318b is 
currently colliding with the intra group medium, 
triggering a large-scale shock and a burst of star formation. 
On the contrary no shocks are observed in BIG and, excluding the H$\alpha$ loop around CGCG97-120, 
no observational evidences of an interaction between this galaxy and BIG are detected.
Looking at Fig. \ref{net}, it is difficult to believe that the association between 
CGCG97-120 and the H$\alpha$ trails is a mere coincidence; however without 
independent observations it is impossible to determine whether or not we are witnessing 
an extremely high velocity interaction as in the SQ.

\section{Conclusions}
The considerable observational material presented in this paper allows us 
to propose an evolutionary 
history for BIG over the last 1-2 Gyr.
Originally BIG was a normal compact group of galaxies 
with a typical velocity dispersion of $\sim$150-200 $\rm km~s^{-1}$, 
composed of at least three galaxies: a massive evolved early type galaxy (CGCG97-125), 
a less massive late type spiral (CGCG97-114) and a gas rich dwarf galaxy 
(the satellite presently merged with CGCG97-125) with a stellar mass $\sim10^{9}\; \rm M_{\odot}$.
Lying in the outskirts of Abell1367 the group has been attracted by the cluster potential and started its 
infall into the cluster core at a speed of $\sim$ 1700 $\rm km~s^{-1}$.
During the journey, its galaxies were perturbed by mutual gravitational interactions, as is often observed in 
compact groups, resulting in star and gas stripping, tidal tails (K2), extragalactic 
compact HII regions (K5 and K1) and tidal dwarfs (DW1, DW2 and DW3). 
Finally the gas rich satellite merged into CGCG97-125 producing 
stellar shells and a burst of star formation.
In the meanwhile tidal interactions also weakened the potential wells of the group galaxies, 
making it easier for ram pressure to strip the galaxies' ISM,
producing the unique H$\alpha$ trails.\\
Galaxy clusters grow by accreting small groups of galaxies 
falling in along large scale filaments. 
The group on which we focused the present investigation might represent
an unique laboratory as it reproduces, at the present epoch, the physical conditions that were likely 
to exist in clusters under formation.   
What can we learn about galaxy/cluster evolution by studying BIG?
The brightest system in BIG, CGCG97-125, appears as a S0 galaxy that 
has recently ($\sim$1.5 Gyr) experienced a minor merging event and 
a consequent burst of star formation,
implying that S0 galaxies can form outside clusters and subsequently join into them.
Groups are the natural sites where gravitational interactions can produce such transformation 
of normal spirals into S0s, a mechanism that has been named \emph{preprocessing}.\\
Tidal interactions within the group's members cannot only
induce morphological transformations in galaxies, but also can
create new systems from stripped material, as shown by the high number of metal rich 
star forming dwarfs/extragalactic HII region detected in BIG.
What is the future of these stripped systems?
If they are dynamically bounded, they might represents the progenitors of 
the population of cluster dwarf galaxies.  Otherwise they will disperse their stars into the intracluster 
medium contributing to the Abell1367's intracluster light. 
The unique H$\alpha$ trails observed in BIG suggest that a considerable amount of 
the ICM enrichment might derive from infalling groups,
as opposed to winds from elliptical galaxies, which are commonly accepted as the major sources of pollution
(e.g. \citealp{delucia04,madau01,mori02}). 
Recent simulations have in fact shown that more of the 10\% of the intracluster medium originated 
from the stripping of gas-rich systems by ram pressure \citep{domainko}. Two more galaxies  
with associated H$\alpha$ trails are indeed observed in the NW part of Abell1367 itself \citep{GAVB01}
emphasizing that a great deal of transformation is taking place in a dynamically young cluster
even at the present cosmological epoch.

\begin{acknowledgements}
We thank the referee, P.A. Duc, for his useful comments which helped us 
to improve and strengthen the paper. 
We wish to thank Samuel Boissier for useful discussions. 
This research has made use of the GOLDMine Database, operated by the Universit\'{a} degli Studi di Milano-Bicocca. 
\end{acknowledgements}

\clearpage

\begin{table*}
\small
\[
\begin{array}{p{0.12\linewidth}rccccccccc}
\hline
\noalign{\smallskip}
Name & \rm R.A. & \rm Dec & \multicolumn{4}{c}{\rm Velocity}  \\
     & \rm (J2000) & \rm (J2000) & \multicolumn{4}{c}{(\rm km~s^{-1})} \\
  &  &  & \rm TNG &\rm ESO-MOS & \rm  Sakai02 & \rm Gavazzi03 \\
\noalign{\smallskip}
\hline
\noalign{\smallskip}
K1            & 11 44 44.18 & 19 48 16.0  &  8422\pm153 & 8265\pm117  &   -  &  8098  \\
DW3~d         & 11 44 45.97 & 19 47 44.4  &    -    & 8564\pm151 &   -  &  -     \\
DW3~e         & 11 44 45.97 & 19 47 41.1  &    -    & 8072\pm124  &   -  &  -     \\
DW3~a         & 11 44 46.43 & 19 47 41.2  &  8490\pm180 &  -         & 8266 &  -     \\
97-114b       & 11 44 46.56 & 19 46 40.3  &    -    & 8656\pm132  & 8504 &  8383  \\
97-114a       & 11 44 47.41 & 19 46 49.8  &    -    & 8763\pm124  &  -   &  8425  \\
K2~a          & 11 44 50.61 & 19 46 05.1  &    -    & 8080\pm140 & 8070 &  8089  \\
K2~b          & 11 44 49.71 & 19 46 04.7  &  8309\pm165 &  -         & -    &   -    \\
DW2~c         & 11 44 51.12 & 19 47 18.7  &  8380\pm188 &  -         & -    &   -    \\
DW2~b         & 11 44 51.17 & 19 47 17.5  &    -    & 8221\pm146 &  -   &  8077  \\
DW2~a         & 11 44 51.67 & 19 47 13.5  &  8253\pm292 &  -         & -    &   -    \\
K5            & 11 44 51.76 & 19 47 52.7  &    -    & 8241\pm112  &  -   &  7995  \\
DW1~b         & 11 44 53.78 & 19 47 31.5  &    -    &  -         & 8070 &   -    \\
DW1~c         & 11 44 54.29 & 19 47 28.6  &  8343\pm223 &  -         & -    &   -    \\
DW1~a         & 11 44 54.64 & 19 47 32.9  &    -    & 8265\pm136  & 8161 &  8067  \\
97-125b       & 11 44 54.89 & 19 46 11.3  &  8261\pm191 & 8396\pm132  & 8170 &  -     \\
K3            & 11 44 55.28 & 19 48 03.3  &  8020\pm212 &  -         &  -   &  -     \\
97-125a       & 11 44 55.99 & 19 46 28.0  &    -        &  -         &  -   &  8330  \\
\noalign{\smallskip}
\hline
\end{array}
\]
\caption{{Redshifts of the galaxies in the BIG group.}
\label{vel}}
\end{table*}
\begin{table*}
\small
\[
\begin{array}{p{0.15\linewidth}rcccccccc}
\hline
\noalign{\smallskip}
Object~~~  & \rm ~~~Tel.~~~  & \rm  C1    & \rm [OII]   & \rm H\beta   & \rm [OIII1] & \rm [OIII2]  & \rm H\alpha & \rm [NII2]  \\
\noalign{\smallskip}
\hline
\noalign{\smallskip}
K1        & ~~~ESO~~~ &   - &  3.68\dagger  &  1.00 &  0.94\dagger &  2.53\dagger &  -   &   -  \\
K1        & ~~~TNG~~~ &  0.00  &  3.49  &  1.00 &  0.95 &  2.57 & 2.86 & 0.58 \\
DW3~a     & ~~~TNG~~~ &  0.02  &  8.26  &  1.00 &  0.31 &  1.06 & 2.86 & 0.59 \\
DW3~d     & ~~~ESO~~~ &  0.00  &  2.01  &  1.00 &  0.60 &  1.34 & 2.86 & 0.69 \\
97-114b   & ~~~ESO~~~ &  0.33  &  4.25  &  1.00 &  0.57 &  1.83 & 2.86 & 0.26 \\
97-114a   & ~~~ESO~~~ &  0.24  &  3.37  &  1.00 &  0.26 &  0.70 & 2.86 & 0.45 \\
CGCG97-114    & ~~~LOI~~~ &  0.75  &  2.61  &  1.00 &  0.21 &  0.36 & 2.86 & 0.64 \\
K2~a      & ~~~ESO~~~ &  0.17  &  4.06  &  1.00 &  0.77 &  2.09 & 2.86 & 0.50 \\
K2~b      & ~~~TNG~~~ &  0.23  &  8.43  &  1.00 &  0.64 &  0.82 & 2.86 & 0.32 \\
DW2~a     & ~~~TNG~~~ &  0.16  &  5.45  &  1.00 &  0.73 &  0.99 & 2.86 & 0.60 \\
DW2~b     & ~~~ESO~~~ &  >0.1  & >5.18  &  1.00\ddagger & <1.00 & <0.99 & 2.86 & <0.61 \\
K5        & ~~~ESO~~~ &  0.56  &    -   &  1.00 &  0.47 &  0.65 & 2.86 & 0.29 \\
DW1~b     & ~~~MMT~~~ &  0.33  &  3.62  &  1.00 &  0.74 &  2.50 & 2.86 & 0.35 \\
DW1~c     & ~~~TNG~~~ &  0.00  &  2.76  &  1.00 &  0.47 &  1.53 & 2.86 & 0.66 \\
DW1~a     & ~~~ESO~~~ &  0.30  &    -   &  1.00 &  0.82 &  2.39 & 2.86 & 0.35 \\
DW1~a     & ~~~MMT~~~ &  0.20  &  3.75  &  1.00 &  0.80 &  2.41 & 2.86 & 0.49 \\
97-125b   & ~~~ESO~~~ &  0.55  &    -   &  1.00 &  0.51 &  1.25 & 2.86 & 0.37 \\
97-125b   & ~~~TNG~~~ &  0.04  &  3.60  &  1.00 &  0.35 &  1.28 & 2.86 & 0.60 \\
CGCG97-125    & ~~~OHP~~~ &  0.88  &  9.23  &  1.00 &  1.06 &  1.97 & 2.86 & 0.84 \\
CGCG97-125    & ~~~TNG~~~ &  0.90  &  6.58  &  1.00 &  1.15 &  1.89 & 2.86 & 1.21 \\
K3        & ~~~TNG~~~ &  0.00  &  2.87  &  1.00 &  0.25 &  0.97 & 2.86 & 0.89 \\
\noalign{\smallskip}
\hline
\end{array}
\]
$\dagger$: Observed values (not corrected for internal extinction)\\
$\ddagger$: H$\beta$ undetected\\
\caption{{Line fluxes, corrected for internal extinction, of the galaxies in the BIG group.}
\label{lines}
}
\end{table*}
\begin{table*}
\small
\[
\begin{array}{p{0.09\linewidth}cccc}
\hline
\noalign{\smallskip}
Name &  r' & \rm H\alpha ~flux & EW(\rm H\alpha+[NII]) & SFR^{a}  \\
     & mag & \rm erg~cm^{-2}~s^{-1}  & \rm \AA &\rm M_{\odot}~yr^{-1} \ \\
\noalign{\smallskip}
\hline
\noalign{\smallskip}
97125 &	13.99   & (1.33\pm0.29) \times 10^{-13} &  27\pm3  & 1.780 \\
97114 &	15.06   & (6.59\pm0.71) \times 10^{-14} &  34\pm5  & 0.880 \\
DW1   & 17.93	& (1.60\pm0.17) \times 10^{-14} & 128\pm15 & 0.113 \\
DW2   &	18.96   & (3.67\pm0.86) \times 10^{-15} &  25\pm7  & 0.023 \\
DW3   &	19.11   & (4.47\pm0.96) \times 10^{-15} &  56\pm15 & 0.027  \\
K1    &	21.46   & (1.35\pm0.20) \times 10^{-15} & 330\pm85 & 0.007 \\
K5    &	22.30   & (1.15\pm0.39) \times 10^{-15} & 650\pm200 & 0.014  \\
97114a & 21.75   & (1.14\pm0.22) \times 10^{-15} & 390\pm120 & 0.008  \\
97114b & 22.00  & (1.42\pm0.38) \times 10^{-15} & 680\pm90 & 0.013  \\
97125b & 21.52   & (1.90\pm0.20) \times 10^{-15} & 370\pm96 & 0.014  \\
\noalign{\smallskip}
\hline
\end{array}
\]
{\tiny a: obtained assuming $\rm SFR=L(H\alpha)/(1.6*10^{41})$ \citep{boselli}, where L(H$\alpha$) is 
the H$\alpha$ luminosity corrected for [NII] contribution 
and internal extinction using values obtained from spectroscopy (see Table \ref{lines}).}
\caption{{Integrated H$\alpha$ properties for some of the star forming systems in BIG.}
\label{tabha}}
\end{table*}
\begin{table*}
\small
\[
\begin{array}{p{0.15\linewidth}rccccccccll}
\hline
\noalign{\smallskip}
Object~~~  & \rm ~~~Tel.~~~  & \multicolumn{7}{c}{\rm 12+log(O/H)}\\
&         & \rm R_{23}^{a}    & \rm R_{23}^{b}   & \rm [NII]/[OII]^{c}   & \rm [NII]/H\alpha^{d} & \rm [OIII]/[NII]^{e}  & \rm Mean & \rm \sigma\\
\noalign{\smallskip}
\hline
\noalign{\smallskip}
K1        & ~~~TNG~~~ & 8.55 & 8.52 & 8.75 & 8.66 & 8.55 & 8.61 & 0.09 \\
DW3~a     & ~~~TNG~~~ & -    & 8.62 & -    & 8.66 & 8.70 & 8.66 & 0.04 \\
DW3~d     & ~~~ESO~~~ & 8.91 & 8.79 & 8.92 & 8.73 & 8.68 & 8.81 &  0.11  \\
97-114b   & ~~~ESO~~~ & 8.59 & 8.52 & -    & 8.30 & 8.48 & 8.47 &  0.13  \\
97-114a   & ~~~ESO~~~ & 8.86 & 8.71 & 8.68 & 8.54 & 8.72 & 8.70 &  0.10  \\
CGCG97-114    & ~~~LOI~~~ & 9.00 & 8.83 & 8.85 & 8.70 & 8.92 & 8.86 &  0.11  \\
K2~a      & ~~~ESO~~~ & 8.56 & 8.51 & 8.65 & 8.59 & 8.56 & 8.57 &  0.05  \\
K2~b      & ~~~TNG~~~ & -    & 8.64 & -    & 8.39 & 8.64 & 8.56 & 0.14 \\
DW2~a     & ~~~TNG~~~ & 8.53 & 8.44 & 8.61 & 8.67 & 8.71 & 8.59 & 0.11 \\
K5        & ~~~ESO~~~ & -    & -    & -    & 8.35 & 8.66 & 8.50 &  0.22  \\
DW1~b     & ~~~MMT~~~ & 8.56 & 8.53 & -    & 8.43 & 8.48 & 8.50 &  0.06  \\
DW1~c     & ~~~MMT~~~ & -    & 8.42 & -    & 8.64 & 8.58 & 8.55 &  0.12  \\
DW1~c     & ~~~TNG~~~ & 8.81 & 8.70 & 8.84 & 8.71 & 8.66 & 8.75 & 0.08 \\
DW1~a     & ~~~ESO~~~ &   -  &   -  & -    & 8.43 & 8.49 & 8.46 &  0.04  \\
DW1~a     & ~~~MMT~~~ & 8.55 & 8.52 & 8.67 & 8.58 & 8.53 & 8.57 &  0.06  \\
CGCG97-125    & ~~~OHP~~~ & -    & 8.73 & -    & 8.82 & 8.65 & 8.73 &  0.08  \\
CGCG97-125    & ~~~TNG~~~ & -    & 8.49 & 8.77 & 8.98 & 8.72 & 8.74 & 0.20 \\
97-125b   & ~~~ESO~~~ & -    & -    & -    & 8.45 & 8.59 & 8.52 &  0.10  \\
97-125b   & ~~~TNG~~~ & 8.76 & 8.64 & 8.75 & 8.67 & 8.67 & 8.69 & 0.05 \\
K3        & ~~~TNG~~~ & 8.89 & 8.75 & 8.90 & 8.84 & 8.78 & 8.83 & 0.07 \\
\noalign{\smallskip}
\hline
\end{array}
\]
a: Zaritsky et al. 1994\\
b: McGaugh 1991\\
c: Kewley \& Dopita 2002\\
d: Van Zee et al. 1998\\
e: Dutil \& Roy 1999
\caption{{Metallicities (12+log(O/H)) of the galaxies in the BIG group.}
\label{metals}}
\end{table*}
\begin{table*}
\small 
\[
\begin{array}{p{0.15\linewidth}rccc}
\hline
\noalign{\smallskip}
Spectrum & Z & Mass & \tau & t \\
       & \rm Z_{\odot} & \log(M/M_{\odot}) & \rm Gyr & \rm Gyr\\
\noalign{\smallskip}
\hline
\noalign{\smallskip}
Nuclear    & 0.04   & 11.01 & \leq 1.0 & 13 \\
Starburst  & 0.04   & 9.27  & 0.8-3.0 & 1.4-2.8\\ 
\noalign{\smallskip}
\hline
\end{array}
\]
\caption{{Best-fitting parameters for the nuclear and starburst component of CGCG97-125.}
\label{fitnucleo}} 
\end{table*}
\clearpage
\setcounter{figure}{12}
\begin{figure*}
\centering
\includegraphics[width=15cm]{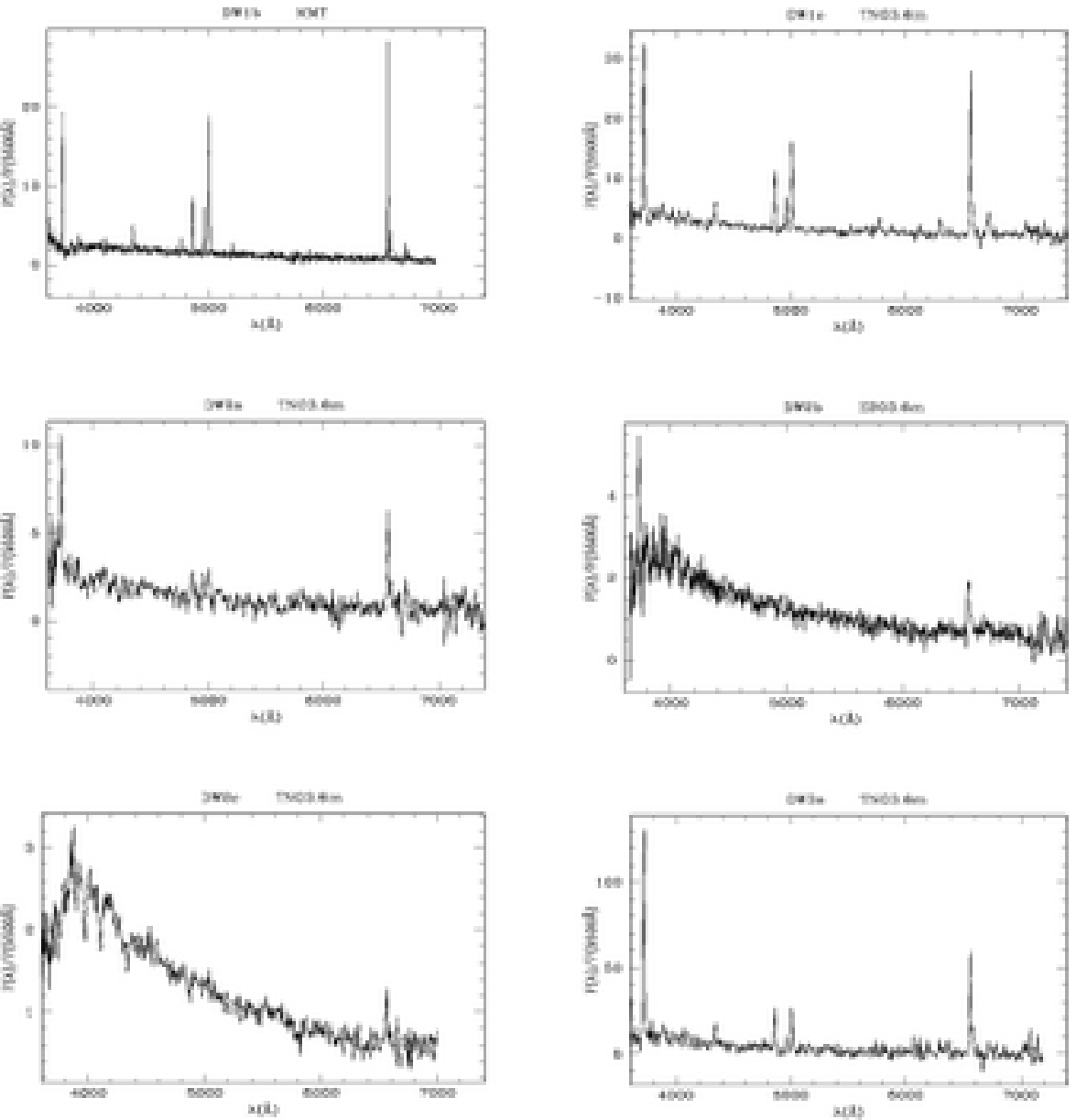}
\small{\caption{Continue.}}
\end{figure*}
\setcounter{figure}{12}
\begin{figure*}
\centering
\includegraphics[width=15cm]{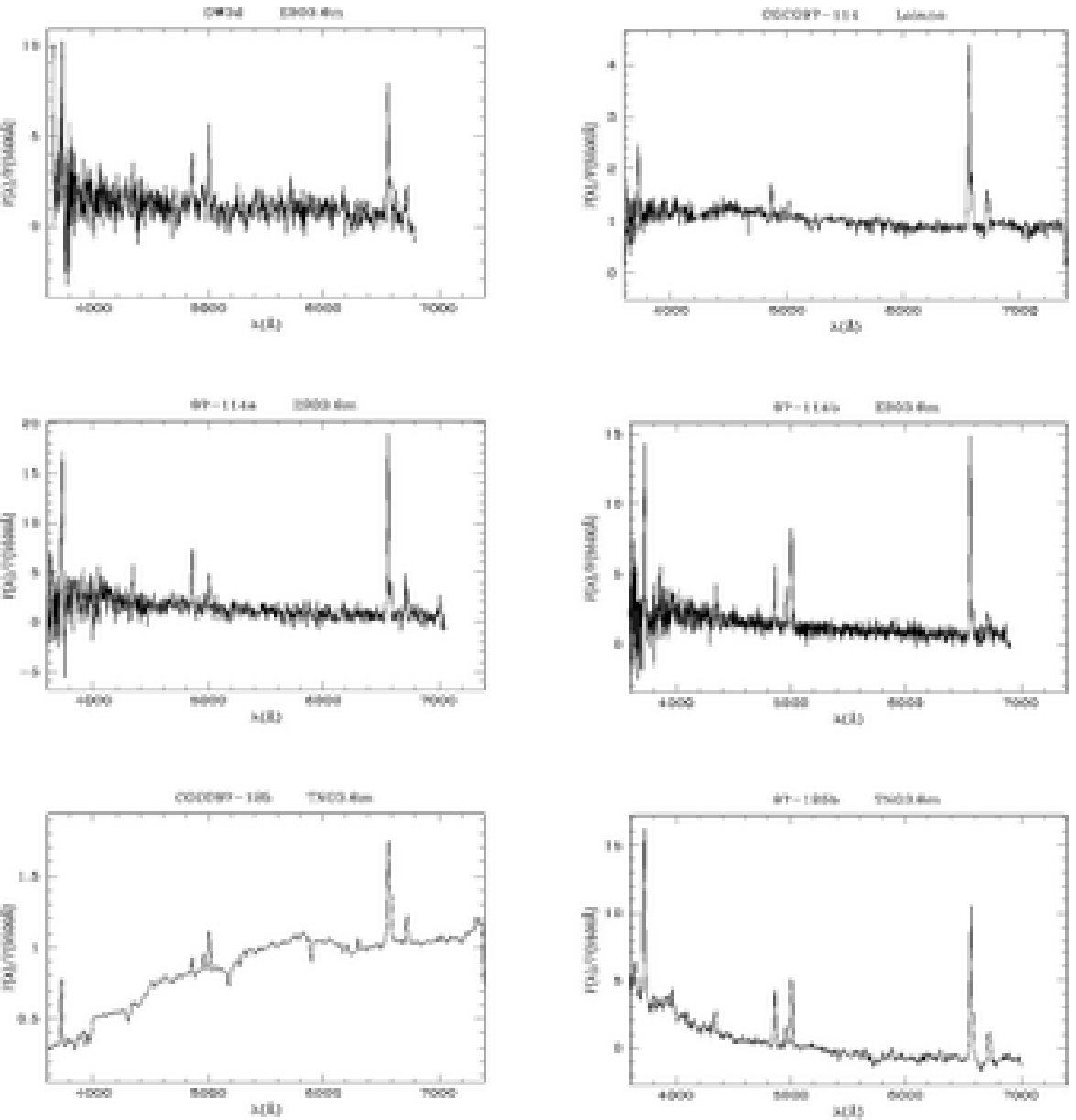}
\small{\caption{Continue.}}
\end{figure*}


\begin{thebibliography}{97}
\expandafter\ifx\csname natexlab\endcsname\relax\def\natexlab#1{#1}\fi

\bibitem[{{Barnes}(2004)}]{barnes04}
{Barnes}, J.~E. 2004, \mnras, 350, 798

\bibitem[{{Boissier} {et~al.}(2003){Boissier}, {Prantzos}, {Boselli}, \&
  {Gavazzi}}]{boissier03}
{Boissier}, S., {Prantzos}, N., {Boselli}, A., \& {Gavazzi}, G. 2003, \mnras,
  346, 1215

\bibitem[{{Boselli} {et~al.}(2005){Boselli}, {Boissier}, {Cortese}, {Gil de
  Paz}, {Buat}, {Iglesias-Paramo}, {Madore}, {Barlow}, {Bianchi}, {Byun},
  {Donas}, {Forster}, {Friedman}, {Heckman}, {Jelinsky}, {Lee}, {Malina},
  {Martin}, {Milliard}, {Morrissey}, {Neff}, {Rich}, {Schiminovich}, {Seibert},
  {Siegmund}, {Small}, {Szalay}, {Welsh}, \& {Wyder}}]{n4438}
{Boselli}, A., {Boissier}, S., {Cortese}, L., {et~al.} 2005, \apjl, 623, L13

\bibitem[{{Boselli} \& {Gavazzi}(2002)}]{bosgav02}
{Boselli}, A. \& {Gavazzi}, G. 2002, \aap, 386, 124

\bibitem[{{Boselli} {et~al.}(2001){Boselli}, {Gavazzi}, {Donas}, \&
  {Scodeggio}}]{boselli}
{Boselli}, A., {Gavazzi}, G., {Donas}, J., \& {Scodeggio}, M. 2001, \aj, 121,
  753

\bibitem[{{Boselli} {et~al.}(2003){Boselli}, {Gavazzi}, \&
  {Sanvito}}]{bosellised}
{Boselli}, A., {Gavazzi}, G., \& {Sanvito}, G. 2003, \aap, 402, 37

\bibitem[{{Boselli} {et~al.}(2002){Boselli}, {Iglesias-P{\' a}ramo},
  {V{\'{\i}}lchez}, \& {Gavazzi}}]{bosha2}
{Boselli}, A., {Iglesias-P{\' a}ramo}, J., {V{\'{\i}}lchez}, J.~M., \&
  {Gavazzi}, G. 2002, \aap, 386, 134

\bibitem[{{Bravo-Alfaro} {et~al.}(2000){Bravo-Alfaro}, {Cayatte}, {van Gorkom},
  \& {Balkowski}}]{hector1}
{Bravo-Alfaro}, H., {Cayatte}, V., {van Gorkom}, J.~H., \& {Balkowski}, C.
  2000, \aj, 119, 580

\bibitem[{{Bresolin} \& {Kennicutt}(1997)}]{bresolin}
{Bresolin}, F. \& {Kennicutt}, R.~C. 1997, \aj, 113, 975

\bibitem[{{Bruzual} \& {Charlot}(2003)}]{BC2003}
{Bruzual}, G. \& {Charlot}, S. 2003, \mnras, 344, 1000

\bibitem[{{Cayatte} {et~al.}(1990){Cayatte}, {van Gorkom}, {Balkowski}, \&
  {Kotanyi}}]{cayatte90}
{Cayatte}, V., {van Gorkom}, J.~H., {Balkowski}, C., \& {Kotanyi}, C. 1990,
  \aj, 100, 604

\bibitem[{{Cortese} {et~al.}(2006){Cortese}, {Boselli}, {Buat}, {Gavazzi},
  {Boissier}, {Gil de Paz}, {Seibert}, {Madore}, \& {Martin}}]{COdust05}
{Cortese}, L., {Boselli}, A., {Buat}, V., {et~al.} 2006, \apj, 637, 242

\bibitem[{{Cortese} {et~al.}(2005){Cortese}, {Boselli}, {Gavazzi},
  {Iglesias-Paramo}, {Madore}, {Barlow}, {Bianchi}, {Byun}, {Donas}, {Forster},
  {Friedman}, {Heckman}, {Jelinsky}, {Lee}, {Malina}, {Martin}, {Milliard},
  {Morrissey}, {Neff}, {Rich}, {Schiminovich}, {Siegmund}, {Small}, {Szalay},
  {Treyer}, {Welsh}, \& {Wyder}}]{CORBGA05}
{Cortese}, L., {Boselli}, A., {Gavazzi}, G., {et~al.} 2005, \apjl, 623, L17

\bibitem[{{Cortese} {et~al.}(2004){Cortese}, {Gavazzi}, {Boselli},
  {Iglesias-Paramo}, \& {Carrasco}}]{COGA04}
{Cortese}, L., {Gavazzi}, G., {Boselli}, A., {Iglesias-Paramo}, J., \&
  {Carrasco}, L. 2004, \aap, 425, 429

\bibitem[{{Cowie} \& {McKee}(1977)}]{COMK77}
{Cowie}, L.~L. \& {McKee}, C.~F. 1977, \apj, 211, 135

\bibitem[{{Cowie} {et~al.}(1981){Cowie}, {McKee}, \& {Ostriker}}]{COMC81}
{Cowie}, L.~L., {McKee}, C.~F., \& {Ostriker}, J.~P. 1981, \apj, 247, 908

\bibitem[{{De Lucia} {et~al.}(2004){De Lucia}, {Kauffmann}, \&
  {White}}]{delucia04}
{De Lucia}, G., {Kauffmann}, G., \& {White}, S.~D.~M. 2004, \mnras, 349, 1101

\bibitem[{{Devine} \& {Bally}(1999)}]{m82cap}
{Devine}, D. \& {Bally}, J. 1999, \apj, 510, 197

\bibitem[{{Domainko} {et~al.}(2005){Domainko}, {Mair}, {Kapferer}, {van
  Kampen}, {Kronberger}, {Schindler}, {Kimeswenger}, {Ruffert}, \&
  {Mangete}}]{domainko}
{Domainko}, W., {Mair}, M., {Kapferer}, W., {et~al.} 2005, astro-ph/0507605

\bibitem[{{Donnelly} {et~al.}(2001){Donnelly}, {Forman}, {Jones}, {Quintana},
  {Ramirez}, {Churazov}, \& {Gilfanov}}]{donnelly01}
{Donnelly}, R.~H., {Forman}, W., {Jones}, C., {et~al.} 2001, \apj, 562, 254

\bibitem[{{Donnelly} {et~al.}(1998){Donnelly}, {Markevitch}, {Forman}, {Jones},
  {David}, {Churazov}, \& {Gilfanov}}]{DON98}
{Donnelly}, R.~H., {Markevitch}, M., {Forman}, W., {et~al.} 1998, \apj, 500,
  138

\bibitem[{{Dressler}(1980)}]{dressler80}
{Dressler}, A. 1980, \apj, 236, 351

\bibitem[{{Dressler}(2004)}]{dress04car}
{Dressler}, A. 2004, in Clusters of Galaxies: Probes of Cosmological Structure
  and Galaxy Evolution, 207

\bibitem[{{Duc} {et~al.}(2000){Duc}, {Brinks}, {Springel}, {Pichardo},
  {Weilbacher}, \& {Mirabel}}]{duc00}
{Duc}, P.-A., {Brinks}, E., {Springel}, V., {et~al.} 2000, \aj, 120, 1238

\bibitem[{{Duc} \& {Mirabel}(1998)}]{duc98}
{Duc}, P.-A. \& {Mirabel}, I.~F. 1998, \aap, 333, 813

\bibitem[{{Duc} \& {Mirabel}(1999)}]{duc99}
{Duc}, P.-A. \& {Mirabel}, I.~F. 1999, in IAU Symp. 186: Galaxy Interactions at
  Low and High Redshift, 61

\bibitem[{{Dutil} \& {Roy}(1999)}]{dutil99}
{Dutil}, Y. \& {Roy}, J. 1999, \apj, 516, 62

\bibitem[{{Ferrari} {et~al.}(2003){Ferrari}, {Maurogordato}, {Cappi}, \&
  {Benoist}}]{ferrari03}
{Ferrari}, C., {Maurogordato}, S., {Cappi}, A., \& {Benoist}, C. 2003, \aap,
  399, 813

\bibitem[{{Forbes} {et~al.}(2003){Forbes}, {Beasley}, {Bekki}, {Brodie}, \&
  {Strader}}]{forbes03}
{Forbes}, D.~A., {Beasley}, M.~A., {Bekki}, K., {Brodie}, J.~P., \& {Strader},
  J. 2003, Science, 301, 1217

\bibitem[{{Franzetti}(2005)}]{paolotesi}
{Franzetti}, P. 2005, Ph.D.~Thesis, Universit\'{a} degli Studi di
  Milano-Bicocca

\bibitem[{{Fujita}(2004)}]{FUJI04}
{Fujita}, Y. 2004, \pasj, 56, 29

\bibitem[{{Gavazzi} {et~al.}(2002{\natexlab{a}}){Gavazzi}, {Bonfanti},
  {Sanvito}, {Boselli}, \& {Scodeggio}}]{gav02}
{Gavazzi}, G., {Bonfanti}, C., {Sanvito}, G., {Boselli}, A., \& {Scodeggio}, M.
  2002{\natexlab{a}}, \apj, 576, 135

\bibitem[{{Gavazzi} {et~al.}(2006){Gavazzi}, {Boselli}, {Cortese}, {Arosio},
  {Gallazzi}, {Pedotti}, \& {Carrasco}}]{ha06}
{Gavazzi}, G., {Boselli}, A., {Cortese}, L., {et~al.} 2006, \aap, in~press

\bibitem[{{Gavazzi} {et~al.}(2003{\natexlab{a}}){Gavazzi}, {Boselli}, {Donati},
  {Franzetti}, \& {Scodeggio}}]{goldmine}
{Gavazzi}, G., {Boselli}, A., {Donati}, A., {Franzetti}, P., \& {Scodeggio}, M.
  2003{\natexlab{a}}, \aap, 400, 451

\bibitem[{{Gavazzi} {et~al.}(2001{\natexlab{a}}){Gavazzi}, {Boselli}, {Mayer},
  {Iglesias-Paramo}, {V{\'{\i}}lchez}, \& {Carrasco}}]{GAVB01}
{Gavazzi}, G., {Boselli}, A., {Mayer}, L., {et~al.} 2001{\natexlab{a}}, \apjl,
  563, L23

\bibitem[{{Gavazzi} {et~al.}(2002{\natexlab{b}}){Gavazzi}, {Boselli},
  {Pedotti}, {Gallazzi}, \& {Carrasco}}]{gavha}
{Gavazzi}, G., {Boselli}, A., {Pedotti}, P., {Gallazzi}, A., \& {Carrasco}, L.
  2002{\natexlab{b}}, \aap, 386, 114

\bibitem[{{Gavazzi} {et~al.}(2002{\natexlab{c}}){Gavazzi}, {Boselli},
  {Pedotti}, {Gallazzi}, \& {Carrasco}}]{haanna}
{Gavazzi}, G., {Boselli}, A., {Pedotti}, P., {Gallazzi}, A., \& {Carrasco}, L.
  2002{\natexlab{c}}, \aap, 396, 449

\bibitem[{{Gavazzi} {et~al.}(1999){Gavazzi}, {Boselli}, {Scodeggio}, {Pierini},
  \& {Belsole}}]{gav99}
{Gavazzi}, G., {Boselli}, A., {Scodeggio}, M., {Pierini}, D., \& {Belsole}, E.
  1999, \mnras, 304, 595

\bibitem[{{Gavazzi} {et~al.}(1998){Gavazzi}, {Catinella}, {Carrasco},
  {Boselli}, \& {Contursi}}]{catinella}
{Gavazzi}, G., {Catinella}, B., {Carrasco}, L., {Boselli}, A., \& {Contursi},
  A. 1998, \aj, 115, 1745

\bibitem[{{Gavazzi} {et~al.}(2003{\natexlab{b}}){Gavazzi}, {Cortese},
  {Boselli}, {Iglesias-Paramo}, {V{\'{\i}}lchez}, \& {Carrasco}}]{GAVC03}
{Gavazzi}, G., {Cortese}, L., {Boselli}, A., {et~al.} 2003{\natexlab{b}}, \apj,
  597, 210

\bibitem[{{Gavazzi} {et~al.}(2001{\natexlab{b}}){Gavazzi}, {Marcelin},
  {Boselli}, {Amram}, {V{\'{\i}}lchez}, {Iglesias-Paramo}, \&
  {Tarenghi}}]{ugc6697}
{Gavazzi}, G., {Marcelin}, M., {Boselli}, A., {et~al.} 2001{\natexlab{b}},
  \aap, 377, 745

\bibitem[{{Gavazzi} {et~al.}(2004){Gavazzi}, {Zaccardo}, {Sanvito}, {Boselli},
  \& {Bonfanti}}]{gavspectra}
{Gavazzi}, G., {Zaccardo}, A., {Sanvito}, G., {Boselli}, A., \& {Bonfanti}, C.
  2004, \aap, 417, 499

\bibitem[{{Gottl{\" o}ber} {et~al.}(2001){Gottl{\" o}ber}, {Klypin}, \&
  {Kravtsov}}]{lambcdm}
{Gottl{\" o}ber}, S., {Klypin}, A., \& {Kravtsov}, A.~V. 2001, \apj, 546, 223

\bibitem[{{Gunn} \& {Gott}(1972)}]{GUNG72}
{Gunn}, J.~E. \& {Gott}, J.~R.~I. 1972, \apj, 176, 1

\bibitem[{{Hameed} {et~al.}(2001){Hameed}, {Blank}, {Young}, \&
  {Devereux}}]{ngc7213}
{Hameed}, S., {Blank}, D.~L., {Young}, L.~M., \& {Devereux}, N. 2001, \apjl,
  546, L97

\bibitem[{{Haynes} {et~al.}(2000){Haynes}, {Jore}, {Barrett}, {Broeils}, \&
  {Murray}}]{haynes00}
{Haynes}, M.~P., {Jore}, K.~P., {Barrett}, E.~A., {Broeils}, A.~H., \&
  {Murray}, B.~M. 2000, \aj, 120, 703

\bibitem[{{Ibata} {et~al.}(2001){Ibata}, {Irwin}, {Lewis}, {Ferguson}, \&
  {Tanvir}}]{ibata01}
{Ibata}, R., {Irwin}, M., {Lewis}, G., {Ferguson}, A.~M.~N., \& {Tanvir}, N.
  2001, \nat, 412, 49

\bibitem[{{Iglesias-P{\' a}ramo} {et~al.}(2002){Iglesias-P{\' a}ramo},
  {Boselli}, {Cortese}, {V{\'{\i}}lchez}, \& {Gavazzi}}]{jorge}
{Iglesias-P{\' a}ramo}, J., {Boselli}, A., {Cortese}, L., {V{\'{\i}}lchez},
  J.~M., \& {Gavazzi}, G. 2002, \aap, 384, 383

\bibitem[{{Iglesias-P{\' a}ramo} \& {V{\'{\i}}lchez}(2001)}]{IGLV01}
{Iglesias-P{\' a}ramo}, J. \& {V{\'{\i}}lchez}, J.~M. 2001, \apj, 550, 204

\bibitem[{{Jog} \& {Solomon}(1992)}]{jog92}
{Jog}, C.~J. \& {Solomon}, P.~M. 1992, \apj, 387, 152

\bibitem[{{Jore} {et~al.}(1996){Jore}, {Broeils}, \& {Haynes}}]{jore}
{Jore}, K.~P., {Broeils}, A.~H., \& {Haynes}, M.~P. 1996, \aj, 112, 438

\bibitem[{{Kauffmann} {et~al.}(2003){Kauffmann}, {Heckman}, {White}, {Charlot},
  {Tremonti}, {Brinchmann}, {Bruzual}, {Peng}, {Seibert}, {Bernardi},
  {Blanton}, {Brinkmann}, {Castander}, {Cs{\' a}bai}, {Fukugita}, {Ivezic},
  {Munn}, {Nichol}, {Padmanabhan}, {Thakar}, {Weinberg}, \& {York}}]{kauff03}
{Kauffmann}, G., {Heckman}, T.~M., {White}, S.~D.~M., {et~al.} 2003, \mnras,
  341, 33

\bibitem[{{Kennicutt} {et~al.}(1989){Kennicutt}, {Keel}, \& {Blaha}}]{kenn89}
{Kennicutt}, R.~C., {Keel}, W.~C., \& {Blaha}, C.~A. 1989, \aj, 97, 1022

\bibitem[{{Kewley} \& {Dopita}(2002)}]{kewley02}
{Kewley}, L.~J. \& {Dopita}, M.~A. 2002, \apjs, 142, 35

\bibitem[{{Kobulnicky} {et~al.}(1999){Kobulnicky}, {Kennicutt}, \&
  {Pizagno}}]{kobulnicky99}
{Kobulnicky}, H.~A., {Kennicutt}, R.~C., \& {Pizagno}, J.~L. 1999, \apj, 514,
  544

\bibitem[{{Kodama} {et~al.}(2001){Kodama}, {Smail}, {Nakata}, {Okamura}, \&
  {Bower}}]{kodama01}
{Kodama}, T., {Smail}, I., {Nakata}, F., {Okamura}, S., \& {Bower}, R.~G. 2001,
  \apjl, 562, L9

\bibitem[{{Kojima} \& {Noguchi}(1997)}]{shell}
{Kojima}, M. \& {Noguchi}, M. 1997, \apj, 481, 132

\bibitem[{{Lequeux} {et~al.}(1979){Lequeux}, {Peimbert}, {Rayo}, {Serrano}, \&
  {Torres-Peimbert}}]{lequex}
{Lequeux}, J., {Peimbert}, M., {Rayo}, J.~F., {Serrano}, A., \&
  {Torres-Peimbert}, S. 1979, \aap, 80, 155

\bibitem[{{Madau} {et~al.}(2001){Madau}, {Ferrara}, \& {Rees}}]{madau01}
{Madau}, P., {Ferrara}, A., \& {Rees}, M.~J. 2001, \apj, 555, 92

\bibitem[{{Maloney} {et~al.}(1996){Maloney}, {Hollenbach}, \&
  {Tielens}}]{malon96}
{Maloney}, P.~R., {Hollenbach}, D.~J., \& {Tielens}, A.~G.~G.~M. 1996, \apj,
  466, 561

\bibitem[{{McGaugh}(1991)}]{mcg91}
{McGaugh}, S.~S. 1991, \apj, 380, 140

\bibitem[{{Mendes de Oliveira} {et~al.}(2004){Mendes de Oliveira}, {Cypriano},
  {Sodr{\' e}}, \& {Balkowski}}]{mendes04}
{Mendes de Oliveira}, C., {Cypriano}, E.~S., {Sodr{\' e}}, L., \& {Balkowski},
  C. 2004, \apjl, 605, L17

\bibitem[{{Mendes de Oliveira} {et~al.}(2001){Mendes de Oliveira}, {Plana},
  {Amram}, {Balkowski}, \& {Bolte}}]{mendes01}
{Mendes de Oliveira}, C., {Plana}, H., {Amram}, P., {Balkowski}, C., \&
  {Bolte}, M. 2001, \aj, 121, 2524

\bibitem[{{Mihos}(2004)}]{mihos04}
{Mihos}, J.~C. 2004, in Clusters of Galaxies: Probes of Cosmological Structure
  and Galaxy Evolution, 278

\bibitem[{{Moore} {et~al.}(1996){Moore}, {Katz}, {Lake}, {Dressler}, \&
  {Oemler}}]{harrassment}
{Moore}, B., {Katz}, N., {Lake}, G., {Dressler}, A., \& {Oemler}, A. 1996,
  \nat, 379, 613

\bibitem[{{Mori} {et~al.}(2002){Mori}, {Ferrara}, \& {Madau}}]{mori02}
{Mori}, M., {Ferrara}, A., \& {Madau}, P. 2002, \apj, 571, 40

\bibitem[{{Morse} {et~al.}(1998){Morse}, {Cecil}, {Wilson}, \&
  {Tsvetanov}}]{ngc5252}
{Morse}, J.~A., {Cecil}, G., {Wilson}, A.~S., \& {Tsvetanov}, Z.~I. 1998, \apj,
  505, 159

\bibitem[{{Mulchaey}(2000)}]{cgroup}
{Mulchaey}, J.~S. 2000, \araa, 38, 289

\bibitem[{{O'Neil}(2004)}]{oneil04}
{O'Neil}, K. 2004, \aj, 128, 2080

\bibitem[{{Oosterloo} \& {van Gorkom}(2005)}]{osterloo05}
{Oosterloo}, T. \& {van Gorkom}, J. 2005, \aap, 437, L19

\bibitem[{{Osterbrock}(1989)}]{osterb89}
{Osterbrock}, D.~E. 1989, {Astrophysics of gaseous nebulae and active galactic
  nuclei} (Research supported by the University of California, John Simon
  Guggenheim Memorial Foundation, University of Minnesota, et al.~Mill Valley,
  CA, University Science Books, 1989, 422 p.)

\bibitem[{{Poggianti} \& {Barbaro}(1997)}]{poggia97}
{Poggianti}, B.~M. \& {Barbaro}, G. 1997, \aap, 325, 1025

\bibitem[{{Poggianti} {et~al.}(2001){Poggianti}, {Bridges}, {Mobasher},
  {Carter}, {Doi}, {Iye}, {Kashikawa}, {Komiyama}, {Okamura}, {Sekiguchi},
  {Shimasaku}, {Yagi}, \& {Yasuda}}]{poggianti01}
{Poggianti}, B.~M., {Bridges}, T.~J., {Mobasher}, B., {et~al.} 2001, \apj, 562,
  689

\bibitem[{{Poggianti} {et~al.}(1999){Poggianti}, {Smail}, {Dressler}, {Couch},
  {Barger}, {Butcher}, {Ellis}, \& {Oemler}}]{poggia99}
{Poggianti}, B.~M., {Smail}, I., {Dressler}, A., {et~al.} 1999, \apj, 518, 576

\bibitem[{{Rubin} {et~al.}(1999){Rubin}, {Waterman}, \& {Kenney}}]{rubin99}
{Rubin}, V.~C., {Waterman}, A.~H., \& {Kenney}, J.~D.~P. 1999, \aj, 118, 236

\bibitem[{{Sakai} {et~al.}(2002){Sakai}, {Kennicutt}, {van der Hulst}, \&
  {Moss}}]{SAKK02}
{Sakai}, S., {Kennicutt}, R.~C., {van der Hulst}, J.~M., \& {Moss}, C. 2002,
  \apj, 578, 842

\bibitem[{{Sakai} {et~al.}(2000){Sakai}, {Mould}, {Hughes}, {Huchra}, {Macri},
  {Kennicutt}, {Gibson}, {Ferrarese}, {Freedman}, {Han}, {Ford}, {Graham},
  {Illingworth}, {Kelson}, {Madore}, {Sebo}, {Silbermann}, \&
  {Stetson}}]{sakai00}
{Sakai}, S., {Mould}, J.~R., {Hughes}, S.~M.~G., {et~al.} 2000, \apj, 529, 698

\bibitem[{{Salpeter}(1955)}]{salpeter}
{Salpeter}, E.~E. 1955, \apj, 121, 161

\bibitem[{{Shang} {et~al.}(1998){Shang}, {Brinks}, {Zheng}, {Chen}, {Burstein},
  {Su}, {Byun}, {Deng}, {Deng}, {Fan}, {Jiang}, {Li}, {Lin}, {Ma}, {Sun},
  {Wills}, {Windhorst}, {Wu}, {Xia}, {Xu}, {Xue}, {Yan}, {Zhou}, {Zhu}, \&
  {Zou}}]{shang98}
{Shang}, Z., {Brinks}, E., {Zheng}, Z., {et~al.} 1998, \apjl, 504, L23+

\bibitem[{{Skillman} {et~al.}(1989){Skillman}, {Kennicutt}, \&
  {Hodge}}]{skillman89}
{Skillman}, E.~D., {Kennicutt}, R.~C., \& {Hodge}, P.~W. 1989, \apj, 347, 875

\bibitem[{{Smail} {et~al.}(1997){Smail}, {Ellis}, {Dressler}, {Couch},
  {Oemler}, {Sharples}, \& {Butcher}}]{smail97}
{Smail}, I., {Ellis}, R.~S., {Dressler}, A., {et~al.} 1997, \apj, 479, 70

\bibitem[{{Sulentic} {et~al.}(2001){Sulentic}, {Rosado}, {Dultzin-Hacyan},
  {Verdes-Montenegro}, {Trinchieri}, {Xu}, \& {Pietsch}}]{sulentic01}
{Sulentic}, J.~W., {Rosado}, M., {Dultzin-Hacyan}, D., {et~al.} 2001, \aj, 122,
  2993

\bibitem[{{Sun} \& {Murray}(2002)}]{SUNM02}
{Sun}, M. \& {Murray}, S.~S. 2002, \apj, 576, 708

\bibitem[{{Treu} {et~al.}(2003){Treu}, {Ellis}, {Kneib}, {Dressler}, {Smail},
  {Czoske}, {Oemler}, \& {Natarajan}}]{treu03}
{Treu}, T., {Ellis}, R.~S., {Kneib}, J.-P., {et~al.} 2003, \apj, 591, 53

\bibitem[{{van Zee} {et~al.}(1998){van Zee}, {Salzer}, {Haynes}, {O'Donoghue},
  \& {Balonek}}]{vanzee98}
{van Zee}, L., {Salzer}, J.~J., {Haynes}, M.~P., {O'Donoghue}, A.~A., \&
  {Balonek}, T.~J. 1998, \aj, 116, 2805

\bibitem[{{Veilleux} {et~al.}(2003){Veilleux}, {Shopbell}, {Rupke},
  {Bland-Hawthorn}, \& {Cecil}}]{veilleux03}
{Veilleux}, S., {Shopbell}, P.~L., {Rupke}, D.~S., {Bland-Hawthorn}, J., \&
  {Cecil}, G. 2003, \aj, 126, 2185

\bibitem[{{Vollmer}(2003)}]{vollmer4654}
{Vollmer}, B. 2003, \aap, 398, 525

\bibitem[{{Vollmer} {et~al.}(2005{\natexlab{a}}){Vollmer}, {Braine}, {Combes},
  \& {Sofue}}]{vollmer05}
{Vollmer}, B., {Braine}, J., {Combes}, F., \& {Sofue}, Y. 2005{\natexlab{a}},
  ArXiv Astrophysics e-prints

\bibitem[{{Vollmer} {et~al.}(2001){Vollmer}, {Cayatte}, {Balkowski}, \&
  {Duschl}}]{vollmer01}
{Vollmer}, B., {Cayatte}, V., {Balkowski}, C., \& {Duschl}, W.~J. 2001, \apj,
  561, 708

\bibitem[{{Vollmer} {et~al.}(2005{\natexlab{b}}){Vollmer}, {Huchtmeier}, \&
  {van Driel}}]{vollmer4254}
{Vollmer}, B., {Huchtmeier}, W., \& {van Driel}, W. 2005{\natexlab{b}}, \aap,
  439, 921

\bibitem[{{Wehner} \& {Gallagher}(2005)}]{wehner05}
{Wehner}, E.~H. \& {Gallagher}, J.~S. 2005, \apjl, 618, L21

\bibitem[{{Williams} {et~al.}(2002){Williams}, {Yun}, \&
  {Verdes-Montenegro}}]{wiliamHI}
{Williams}, B.~A., {Yun}, M.~S., \& {Verdes-Montenegro}, L. 2002, \aj, 123,
  2417

\bibitem[{{Xu} {et~al.}(1999){Xu}, {Sulentic}, \& {Tuffs}}]{xu99}
{Xu}, C., {Sulentic}, J.~W., \& {Tuffs}, R. 1999, \apj, 512, 178

\bibitem[{{Xu} {et~al.}(2005){Xu}, {Iglesias-P{\'a}ramo}, {Burgarella}, {Rich},
  {Neff}, {Lauger}, {Barlow}, {Bianchi}, {Byun}, {Forster}, {Friedman},
  {Heckman}, {Jelinsky}, {Lee}, {Madore}, {Malina}, {Martin}, {Milliard},
  {Morrissey}, {Schiminovich}, {Siegmund}, {Small}, {Szalay}, {Welsh}, \&
  {Wyder}}]{xu05}
{Xu}, C.~K., {Iglesias-P{\'a}ramo}, J., {Burgarella}, D., {et~al.} 2005, \apjl,
  619, L95

\bibitem[{{Xu} {et~al.}(2003){Xu}, {Lu}, {Condon}, {Dopita}, \& {Tuffs}}]{xu03}
{Xu}, C.~K., {Lu}, N., {Condon}, J.~J., {Dopita}, M., \& {Tuffs}, R.~J. 2003,
  \apj, 595, 665

\bibitem[{{Zaritsky} {et~al.}(1994){Zaritsky}, {Kennicutt}, \&
  {Huchra}}]{zaritsky94}
{Zaritsky}, D., {Kennicutt}, R.~C., \& {Huchra}, J.~P. 1994, \apj, 420, 87

\bibitem[{{Zwicky} {et~al.}(1961){Zwicky}, {Herzog}, \& {Wild}}]{ZWHE61}
{Zwicky}, F., {Herzog}, E., \& {Wild}, P. 1961, {Catalogue of galaxies and of
  clusters of galaxies} (Pasadena: California Institute of Technology (CIT))

\end{thebibliography}
\end{document}